\documentclass[]{article} % twoside
\usepackage[]{amsmath}
\usepackage{bm}
\usepackage[braket, qm]{qcircuit}
\usepackage{graphicx}
\usepackage{float}
\usepackage{subcaption}
\usepackage{multirow}
\usepackage{multicol}
\usepackage{threeparttable}
\usepackage[title]{appendix}
\usepackage{overpic}
\usepackage[a4paper, total={7in, 10in}]{geometry}
\usepackage{authblk}
\usepackage[utf8]{inputenc}
\usepackage{balance}
\usepackage{times}
\usepackage{dsfont}
\usepackage[T1]{fontenc}
\usepackage{marvosym}
\usepackage[]{ifsym}

\usepackage{booktabs}
\usepackage{collcell}
\newcommand*{\movedown}[1]{%
  \smash{\raisebox{-2.5ex}{#1}}}
\newcolumntype{q}{>{\collectcell\movedown}r<{\endcollectcell}}

\usepackage[marginal]{footmisc}
\usepackage[
backend=biber,
style=nature,
sorting=none,
isbn=false,
date=year,
backref=true,
eprint=false,
]{biblatex}

\usepackage{hyperref}
\hypersetup{
colorlinks=true,
linkcolor=blue,
filecolor=blue,
urlcolor=red,
citecolor=blue,
}
\usepackage{etoolbox}

\usepackage{flushend,cuted}

\title{Variational quantum eigensolver with linear depth problem-inspired ansatz for solving portfolio optimization in finance\vspace{0em}}
\date{}
\author[1,2]{Shengbin Wang}
\author[2]{Peng Wang}
\author[3]{Guihui Li}
\author[1]{Shubin Zhao}
\author[1]{Dongyi Zhao}
\author[1]{Jing Wang}
\author[1]{Yuan Fang}
\author[1]{Menghan Dou}
\author[4]{Yongjian Gu}
\author[2,5,*]{Yu-Chun Wu}
\author[1,2,5,*]{Guo-Ping Guo}

\affil[1]{\small Origin Quantum Computing Company Limited, Hefei, Anhui 230026, People’s Republic of China}
\affil[2]{CAS Key Laboratory of Quantum Information, University of Science and Technology of China, Hefei, Anhui 230026, People’s Republic of China}
\affil[3]{Intelligent Information Sensing and Processing Lab, College of Electronic Engineering, Ocean University of China, Qingdao, Shandong 266000, People’s Republic of China}
\affil[4]{College of Physics and Optoelectronic Engineering, Ocean University of China, Qingdao, Shandong 266100, People’s Republic of China}
\affil[5]{Institute of Artificial Intelligence, Hefei Comprehensive National Science Center, Hefei, Anhui 230088, People’s Republic of China}

\begin{document}
\maketitle
\thispagestyle{empty} %当前页不显示页码

\begin{abstract}
\noindent Great efforts have been dedicated in recent years to explore practical applications for noisy intermediate-scale quantum (NISQ) computers, which is a fundamental and challenging problem in quantum computing. As one of the most promising methods, the variational quantum eigensolver (VQE) has been extensively studied. In this paper, VQE is applied to solve portfolio optimization problems in finance by designing two hardware-efficient Dicke state ansatze that reach a maximum of $2n$ two-qubit gate depth and $\frac{n^2}{4}$ parameters, with $n$ being the number of qubits used. Both ansatze are partitioning-friendly, allowing for the proposal of a highly scalable quantum/classical hybrid distributed computing (HDC) scheme. Combining simultaneous sampling, problem-specific measurement error mitigation, and fragment reuse techniques, we successfully implement the HDC experiments on a superconducting quantum computer with up to 55 qubits. The simulation and experimental results illustrate that the restricted expressibility of the ansatze, induced by the small number of parameters and limited entanglement, is advantageous for solving classical optimization problems with the cost function of the conditional value-at-risk (CVaR) for the NISQ era and beyond. Furthermore, the HDC scheme shows great potential for achieving quantum advantage in the NISQ era. We hope that the heuristic idea presented in this paper can motivate fruitful investigations in current and future quantum computing paradigms.

\end{abstract}

\renewcommand{\thefootnote}{}
\footnotetext{\footnotesize $^{*}$Author to whom any correspondence should be addressed. 

\noindent \textbf{Email}: wuyuchun@ustc.edu.cn; gpguo@ustc.edu.cn}

\section{Introduction}
\label{sec:intro}
Quantum computing can provide astonishing speedups in certain applications, such as simulating quantum systems \cite{Feynman1982}, factoring numbers \cite{Shor1994}, searching unstructured databases \cite{Grover1997}, and solving linear systems of equations \cite{HarrowHassidimLloyd2009}, compared to its classical counterpart. These advantages arise from the peculiar properties of quantum entanglement, superposition and interference. Mixed blessings, these properties also seriously impede the design and manufacture of large-scale fault-tolerant quantum computers that can fully realize the speedup potential of quantum computing \cite{NielsenChuang2016}. Whereupon, the currently available noisy intermediate-scale quantum (NISQ) computer, with no more than hundreds of noisy qubits, short coherence time, and limited connectivity, is the only quantum platform available to us in the coming few years, and possibly even longer \cite{Preskill2018,Leymann2020}. And the widely held focus is on finding the "killer apps" for NISQ computers, which is a notoriously difficult problem \cite{Bharti2022,Lau2022}.

Variational quantum algorithms (VQAs) \cite{Cerezo2021}, developed for efficient implementation on both quantum and classical computers in a hybrid manner, significantly enhance the prospect of practical applications that may offer quantum advantages in the NISQ era. The prominent algorithms are the variational quantum eigensolver (VQE) \cite{PeruzzoVQE2014} and quantum approximate optimization algorithm (QAOA) \cite{FarhiQAOA2014}. In recent years, significant efforts, particularly in the fields of quantum chemistry, quantum machine learning, and combinatorial optimization \cite{Cerezo2021,Kandala2017,Moll2018,McArdle2020,Tilly2022,OBrien2022,Fedorov2022,Biamonte2017,Cong2019,Li2022,Lamata2023,Nannicini2019,Liu2022,Palackal2023}, have been made with VQAs to advance the practicality of NISQ computers. For combinatorial optimization, which has already been extensively applied in many areas of business and science, the studies in the quantum computing field mainly focus on maximum cut, number partitioning, and portfolio optimization, etc \cite{Wang2018,Nannicini2019,BarkoutsosCVaR2020,Yu2022,Amaro2022,Liu2022,ZhuQAOAmaxcut2022}. Compared to QAOA, which encodes the total energy of the problem through the cost Hamiltonian that requires to be translated into a unitary operator, resulting in a much deeper quantum circuit, VQE with the unique quantum part, i.e. the ansatz, and classically optimized expectation of the total energy is more appealing. This comes from the consensus that NISQ computers are limited to executing lower-depth circuits and hence should focus on handling the classically intractable parts of the computation process as efficiently as possible.

The ansatz, a structurally parameterized quantum circuit (PQC) $U(\bm{\theta})$ with parameters $\bm{\theta}$, is the key ingredient of the VQE algorithms. The trial quantum states $\vert\psi(\bm{\theta})\rangle$ prepared by the ansatz is subsequently iteratively sampled and optimized to classically evaluate the minimum energy of the Hamiltonian $H$ of the problem, i.e. $\operatorname{min}_{\bm{\theta}} \langle \psi(\bm{\theta})\vert H \vert \psi(\bm{\theta})\rangle$, along with the corresponding eigenstate, upon the corresponding parameters \bm{$\theta$} \cite{BenedettiPQC2019,Herasymenko2021}. In other words, as the unique quantum part, ansatz determines the output fidelity of NISQ computers and the potential advantage that the VQE algorithms can achieve. Unsuitable ansatz deployed for specific problems trivially leads to bad performance. There are two main categories of ansatze in the context of this paper: hardware-efficient (HE) ansatz and problem-inspired ansatz \cite{Cerezo2021,Fedorov2022}. The versatility HE ansatz that is designed based on considering the restrictive conditions of NISQ computers can mitigate the effect of hardware noise to some extent. That is, HE ansatz is problem-agnostic and can be a circuit with very low depth, even a product state composed of one layer of single-qubit gates. These advantages have garnered much attention, resulting in a flurry of promising results on solving small-scale problems of interest \cite{PeruzzoVQE2014,Kandala2017,BarkoutsosCVaR2020}, and the current combinatorial optimization invests are mainly focusing on this type of ansatz. However, for most of the underlying problems, the problem-agnostic nature of this type of ansatz may lead to undesired issues such as barren plateau (BP), local minima, and poor expressibility \cite{CerezoBP2021,MarreroBP2021,Leone2022,Holmes2022}, particularly when the number of required qubits to span the search space becomes slightly larger. In contrast, a problem-inspired ansatz has access to some prior knowledge of the problem, which can span a more accurate search space. Hence the issues  of BP and local minima can be largely alleviated \cite{BhartiPIA2021}. Unfortunately, this type of ansatz, e.g. the extensively researched unitary coupled cluster (UCC) ansatz and heuristic ansatz in quantum chemistry \cite{Cerezo2021,Gard2020}, typically results in dense entanglement among qubits and leads to much deeper circuits, which is prohibitive for the current NISQ computers to execute efficiently even on solving small-scale problems.

Clearly, viable ansatze for efficiently solving problems of interest on NISQ computers should possess both hardware-efficient and problem-inspired properties simultaneously. That is, this type of ansatz must be a shallow circuit with nearest-neighbor coupling which unfolds a feasible search subspace that encompasses the optimal solution (or the near-optimal solutions) while avoiding, if possible, the generation of irrelevant, even distant, states. In consequence, the key question at present turns to whether such a viable ansatz can be devised, at least for solving specific problems. Besides, another issue that arises is how to mitigate the errors induced by the width (number of qubits) of the ansatz when solving larger-scale problems.

In this paper, we address these questions by introducing two ansatze that can efficiently prepare arbitrary Dicke states \cite{Dicke1954}, which are well-suited for NISQ computers, not only for solving portfolio optimization problems in finance \cite{Orús2019,Herman2022}. Our ansatze are unified frameworks that span the feasible subspace and have a linear circuit depth with a constant factor of at most 2 on $CNOT$ gates, which are applied only between adjacent qubits. The worst-case asymptotic complexity on parameters  \bm{$\theta$} is squared with a factor not exceeding $\frac{1}{4}$. The above achievements make our ansatze the most advanced methods for preparing Dicke states that can be executed on NISQ computers with linear-nearest-neighbor coupling currently and answer the first question in the affirmative. Nevertheless, to guarantee that NISQ computers can be deployed to efficiently solve some larger-scale problems, addressing the second issue is an equally crucial task. Fortunately, the building blocks of our ansatze construct quantum circuits using staircase structure, similar to the structure of the Matrix product state, that is more convenient for us to embed circuit cutting \cite{Peng2020,Ying2023} or qubit reuse \cite{Hua2022} techniques, or their combination, to improve the performance of NISQ computers in terms of circuit width dimension. Further progress has been achieved by proposing a quantum/classical hybrid distributed computing (HDC) scheme. This scheme splits classically the ansatz into several subcircuits with a staircase structure first, and then partitions each subcircuit into applicable fragments using the circuit cutting technique or executes each subcircuit directly using the qubit reuse technique. In the circuit cutting case, the total number of qubits required to be cut in each circuit is greatly reduced. The $X$, $Y$, $Z$ ``measure-and-prepare'' channels are eliminated based on the symmetry of Dicke states. Compared to the general sampling complexity that grows exponentially with the total number of cuts, the sampling complexity of our scheme is unrelated to the total number of cuts. We also significantly reduce the number of submissions of the circuits to the hardware. This preserves completely the advantage of NISQ computers on solving portfolio optimization, not only in the scenarios of selecting a small number of assets. In the qubit reuse case, the number of qubits used to execute each circuit can be reduced to merely 2. We provide detailed comparisons of our scheme to the commonly used ansatze through complexity analysis, numerical simulations, and hardware experiments, verifying the superiority of our methods over the state-of-the-art ones.

The paper is organized as follows. In section \ref{sec:port op}, an overview of the binary integer portfolio optimization problem is provided. The design inspiration, theoretical insights, single-layer structure rationality, detailed optimal circuit and complexity comparisons of the proposed ansatze are introduced in section \ref{sec:vqe ansatz}. In section \ref{sec:experi}, we provide the numerical simulations and hardware experiments of the proposed algorithms, as well as the quantum/classical hybrid distributed computing scheme, simultaneous sampling method, problem-specific measurement error mitigation and fragment reuse technique. Finally, we conclude our work and discuss some interesting problems in section \ref{sec:disc&concl}. 

\section{Portfolio optimization}
\label{sec:port op}
As one of the most common optimization problems in finance, portfolio optimization involves selecting from an asset pool a set of optimally allocated assets that achieve the expected return with minimum financial risk, or maximize the expected return for a given level of risk \cite{Orús2019,Herman2022}. In this paper, we focus on binary integer portfolio optimization based on mean-variance analysis (modern portfolio theory) because its binary variable representation aligns with the prevalent two-level quantum computers. The risk-return optimization can be modeled as
\begin{equation}
\begin{split}
    \operatorname{min}&_{\bm{x}}\ {q\bm{x}^TA\bm{x}}-\bm{\mu}^T\bm{x},\\
    s.t.& \ \xi =\bm{\Pi}^T\bm{x}, \label{eq:mean-variance}
\end{split}
\end{equation}
where $\bm{x}=(x_1,x_2,\ldots,x_n)^T$ with $x_i\in\{0, 1\}$ represents the asset selection vector, i.e. the \emph{i}th asset is selected when boolean variable $x_i=1$ or not when $x_i=0$. $A$ is the $n\times n$ real covariance matrix between assets. The risk level $q>0$ represents the investor's risk tolerance, and the vector $\bm{\mu}$ contains the expected returns of the assets. The constraint is given by $\xi =\bm{\Pi}^T\bm{x}$, where $\xi$ is the budget and $\bm{\Pi}$ is the asset price vector, which can be simplified to the all-ones vector, $\mathds{1}$, in the present scenario.

There are two ways to map the problem to a quadratic unconstrained binary optimization (QUBO) form suitable for quantum computing \cite{Hodson2019}. The general way is known as the soft constraint form, which encodes the constraint to a penalty, $(\xi -\mathds{1}^T\bm{x})^2$, absorbed into the cost function with a penalty scaling coefficient. That is, by transforming variable $x_i\in\{0,1\}$ to $z_i\in\{-1,+1\}$ based on the relation $z_i=1-2x_i$, the binary model is converted to a spin model
\begin{equation}
\operatorname{min}_{\bm{z}}\ {q^{'}\bm{z}^TA^{'}\bm{z}}-\bm{\mu}^{{'}T}\bm{z}+\{\beta(\xi^{'} -\mathds{1}^T\bm{z})^2\}_{soft}, \label{eq:mean-variance spin}
\end{equation}
where $\beta$ is the penalty scaling coefficient. The problem described by equation \eqref{eq:mean-variance spin} can be easily translated into a diagonal Hamiltonian, whose ground state encodes the optimal solution of the optimization problem \eqref{eq:mean-variance} by replacing $z_i$ with $\sigma^i_Z$. Finally, the hardware-efficient (HE) ansatz-based VQE algorithm can be applied to estimate the ground state. However, the landscape spanned by equation \eqref{eq:mean-variance spin} with the HE ansatz is too large and flat and contains too many local minima, which impact severely the optimization of the parameters and, consequently, hinder the convergence of the solution for solving larger-scale problems.

Contrastingly, the hard constraint form encodes the constraint directly into the structure of the problem-inspired ansatz. That is, any vector $\bm{x}$ that does not conform to the constraint $\xi =\mathds{1}^T\bm{x}$ is excluded from the space spanned by the ansatz. And equation \eqref{eq:mean-variance spin} is reduced to
\begin{equation}
\operatorname{min}_{\bm{z}}\ {q^{'}\bm{z}^TA^{'}\bm{z}}-\bm{\mu}^{{'}T}\bm{z}. \label{eq:mean-variance spin no constraint}
\end{equation}
The elimination of the penalty term simplifies the evaluation of $\operatorname{min}_{\bm{\theta}} \langle \psi(\bm{\theta})\vert H \vert \psi(\bm{\theta})\rangle$ on a much smaller landscape.

\section{Variational Dicke state ansatz}
\label{sec:vqe ansatz}
An ansatz is a dynamic quantum state preparation circuit that depends on some parameters optimized through the minimization or maxmization of a problem's cost function. The form of the ansatz determines the form of the cost Hamiltonian and the training behavior of the parameters. In this section, two universal ansatze for preparing arbitrary Dicke states $\vert D^n_k\rangle$ with variable amplitudes as a function of the parameters \bm{$\theta$} are presented. In the present scenario, Dicke state $\vert D^n_k\rangle$ is an \emph{n}-qubit highly entangled state composed of all $\binom{n}{k}$ basis states that form the search space for combinatorial optimization quantum algorithms \cite{Nannicini2019,Liu2022,Palackal2023,Mukherjee2020,Bärtschi2022}. In portfolio optimization, \emph{n} denotes the number of assets that the asset pool can provide and $k\in[1,n-1]$ is the number of assets selected from the pool which is equivalent to the budget $\xi$ \cite{Herman2022}. The applications of Dicke states also involve quantum game \cite{Ozdemir2007}, quantum metrology \cite{Pezze2018}, quantum networks \cite{Miguel2020}, and so on.
\begin{figure}[ht]
    \centerline{
     \Qcircuit @C = 0.7em @R = 0.4em  @!R{
      &\lstick{\vert0\rangle} &\gate{X} &\ctrl{1}   &\targ     &\qw        &\qw       &\qw        &\qw       &\qw        &\qw       &\qw   \\\
      &\lstick{\vert0\rangle} &\qw      &\gate{R_y} &\ctrl{-1} &\ctrl{1}   &\targ     &\qw        &\qw       &\qw        &\qw       &\qw   \\\
      &\lstick{\vert0\rangle} &\qw      &\qw        &\qw       &\gate{R_y} &\ctrl{-1} &\ctrl{1}   &\targ     &\qw        &\qw       &\qw   \\\
      &\lstick{\vert0\rangle} &\qw      &\qw        &\qw       &\qw        &\qw       &\gate{R_y} &\ctrl{-1} &\ctrl{1}   &\targ     &\qw   \\\
      &\lstick{\vert0\rangle} &\qw      &\qw        &\qw       &\qw        &\qw       &\qw        &\qw       &\gate{R_y} &\ctrl{-1} &\qw   \gategroup{1}{4}{5}{11}{2em}{--}
     }}
    \caption{An illustrative staircase-structure circuit for preparing state $\vert W_5\rangle$.}
    \label{fig:Wstate}
\end{figure}
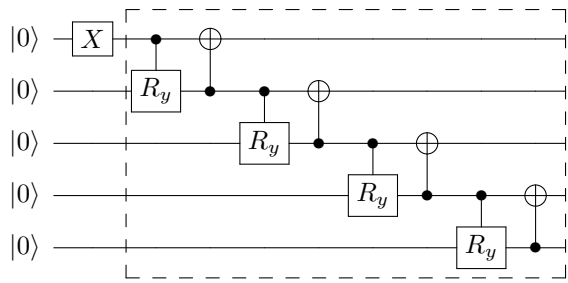

Both ansatze are inspired by the fundamental observation that all the useful information is implicitly stored in the unitary $U_n$, composed of the building blocks ``Controlled-$R_y$ and \emph{CNOT}'' shown in the dashed box of figure \ref{fig:Wstate} for preparing \emph{W} state \cite{Cruz2019}. Hence, we can easily extract the target states by preparing a linear combination of the appropriate columns of \emph{$U_n$}, somewhat similar to the idea of the LCU-based quantum state preparation algorithms \cite{Wang2021a}. In this paper we refer to the circuit structure similar to the one in the dashed box as the `staircase structure,' denoted as \emph{$U_n$}, which is the critical circuit structure of our ansatze. We analyze the theoretical support for both ansatze based on this observation, see appendix \ref{appen: theoretical support} for details. For an arbitrary basis state of $\vert D^n_k\rangle$, there always exists a basis state with reverse symmetry to it, e.g. 11000 with 00011, 01010 with itself. Therefore, the qubit at the top can be regarded as either the lowest qubit or the highest qubit. In the following context, we fix the top qubit to be the highest qubit.

\begin{figure}[ht]
    \centerline{
    \Qcircuit @C = 0.7em @R = 0.4em  @!R{
     &\lstick{\vert0\rangle} &\gate{X}  &\targ     &\ctrl{1}   &\targ      &\qw       &\qw        &\qw        &\qw       &\qw        &\qw       &\qw \\\
     &\lstick{\vert0\rangle} &\qw       &\ctrl{-1} &\gate{R_y} &\ctrl{-1}  &\targ     &\ctrl{1}   &\targ      &\qw       &\qw        &\qw       &\qw \\\
     &\lstick{\vert0\rangle} &\gate{X}  &\targ     &\ctrl{1}   &\targ      &\ctrl{-1} &\gate{R_y} &\ctrl{-1}  &\targ     &\ctrl{1}   &\targ     &\qw \\\
     &\lstick{\vert0\rangle} &\qw       &\ctrl{-1} &\gate{R_y} &\ctrl{-1}  &\targ     &\ctrl{1}   &\targ      &\ctrl{-1} &\gate{R_y} &\ctrl{-1} &\qw \\\
     &\lstick{\vert0\rangle} &\qw       &\qw       &\qw        &\qw        &\ctrl{-1} &\gate{R_y} &\ctrl{-1}  &\qw       &\qw        &\qw       &\qw \\\ \gategroup{1}{4}{2}{6}{1.2em}{--}
    }}
    \caption{Illustration of CCC ansatz for preparing $\vert D^5_2\rangle$. A 3C block is indicated in the dashed box. The variational parameters of $R_y$ gates are omitted for brevity.}
    \label{fig:first ansatz}
\end{figure}
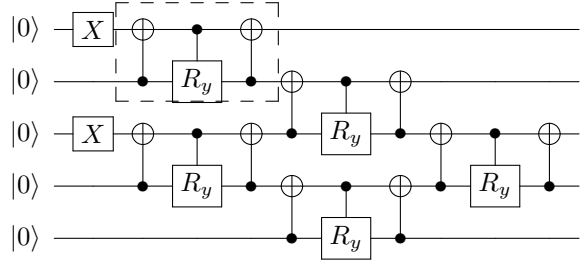
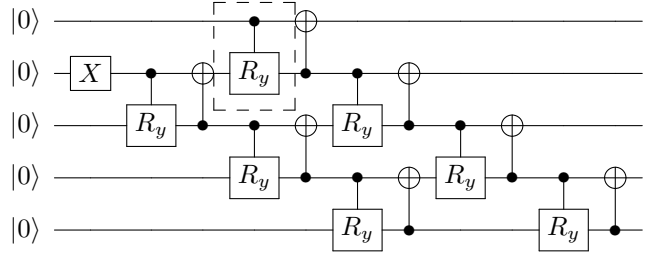
\begin{figure}[ht]
    \centerline{
    \Qcircuit @C = 0.6em @R = 0.4em  @!R{
     &\lstick{\vert0\rangle} &\qw       &\qw        &\qw       &\ctrl{1}   &\targ     &\qw         &\qw        &\qw        &\qw       &\qw        &\qw       &\qw \\\
     &\lstick{\vert0\rangle} &\gate{X}  &\ctrl{1}   &\targ     &\gate{R_y} &\ctrl{-1} &\ctrl{1}    &\targ      &\qw        &\qw       &\qw        &\qw       &\qw \\\
     &\lstick{\vert0\rangle} &\qw       &\gate{R_y} &\ctrl{-1} &\ctrl{1}   &\targ     &\gate{R_y}  &\ctrl{-1}  &\ctrl{1}   &\targ     &\qw        &\qw       &\qw \\\
     &\lstick{\vert0\rangle} &\qw       &\qw        &\qw       &\gate{R_y} &\ctrl{-1} &\ctrl{1}    &\targ      &\gate{R_y} &\ctrl{-1} &\ctrl{1}   &\targ     &\qw \\\
     &\lstick{\vert0\rangle} &\qw       &\qw        &\qw       &\qw        &\qw       &\gate{R_y}  &\ctrl{-1}  &\qw        &\qw       &\gate{R_y} &\ctrl{-1} &\qw \\\ \gategroup{1}{6}{2}{6}{1.2em}{--}
    }}
    \caption{Illustration of CC ansatz for preparing $\vert D^5_2\rangle$. The inoperative controlled-$R_y$ in the dashed box is shown for structure completeness.}
    \label{fig:second ansatz}
\end{figure}

For combinatorial optimization problems, the solution is a single vector (binary string) that is fundamentally different from the ground state composed of multiple vectors for chemical systems. Therefore, the extra degrees of freedom, which should be preserved in solving certain chemistry problems, can be removed in combinatorial optimization scenario. Then based on the observation of \emph{W} state, we construct the staircase unitary $U_n$ using the simplified ``\emph{A}'' gates \cite{Gard2020} which is composed of multiple ``\emph{CNOT}, Controlled-$R_y$, \emph{CNOT}'' (3C) blocks, see figure \ref{fig:first ansatz} for an example. Unexpectedly, the unitary contains all the information about Dicke states $\vert D^n_k\rangle$, which are distributed across different columns, see  appendix \ref{appen: theoretical support} for detailed analysis. Motivated by this observation, the first ansatz is constructed with $k$ layers of staircases. Each layer is composed of $(n-k-\lfloor i/2 \rfloor)$ 3C blocks, hence is called CCC ansatz, starting with an \emph{NOT (X)} gate deployed at the \emph{i}th qubit where \emph{i} equals $2k-1$, $2k-3$, ..., $1$, when $k\leq\lfloor n/2 \rfloor$, see figure \ref{fig:first ansatz}. For the cases of $k> \lfloor n/2 \rfloor$, we use the equivalence $\binom{n}{k}=\binom{n}{n-k}$ to simplify the circuit design, i.e. using the circuit for preparing $\vert D^n_{n-k}\rangle$ followed by the unitary $\otimes^n_{j=1}X_j$ to prepare $\vert D^n_k\rangle$. At last we optimize each 3C block to its optimal circuit that contains only $2$ \emph{CNOT} gates to reduce the circuit depth further, see appendix \ref{appen: optimal theory} for detailed analysis. 

% \end{multicols}
\begin{table}[ht]
    \setlength{\abovecaptionskip}{0.1cm}
    \setlength{\belowcaptionskip}{0.1cm}
    \caption{\small Comparison of Dicke states preparation schemes.}
    \label{tab:Dicke states schemes}
    \centering
    \begin{threeparttable}
    \scalebox{0.7}{
    \renewcommand\arraystretch{2}{
    \begin{tabular}{c c c c c}
    \hline\hline
    Method                           &Depth of \emph{CNOT}s &Number of \emph{CNOT}s &Number of $\theta$s                  &Topology   \\ \hline
    Mukherjee et. al. \cite{Mukherjee2020}                &$O(nk)$                            &$5nk-5k^2$              &$\backslash$                        &all-to-all \\ \hline
    \multirow{2}{*}{Bärtschi et. al. \cite{Bärtschi2022}} &$O(klog\frac{n}{k})$               &$O(nk)$                 &$\backslash$                        &all-to-all \\ 
                                     &$O(k\sqrt{\frac{n}{k}})$           &$O(nk)$                 &$\backslash$                        &Grid       \\ \hline
    \multirow{2}{*}{Ours}            &$2(n-k)$                           &$2nk-3k^2$              &$nk-\frac{3k^2}{2}$               &LNN\tnote{1}        \\ 
                                     &$2n$                               &$nk-\frac{k^2}{2}$      &$\frac{n(k+1)}{2}-\frac{k^2}{4}$  &LNN        \\ \hline\hline
    \multicolumn{5}{l}{
    \begin{minipage}{8cm}
     $^1$ Linear-Nearest-Neighbor connectivity.
    \end{minipage}
    }\\
    \vspace{-3em}
    \end{tabular}}}
    \end{threeparttable}
\end{table}
% \begin{multicols}{2}

The second ansatz proposed here is inspired directly by the circuit for preparing \emph{W} state \cite{Cruz2019}, the \emph{n}-qubit case of which is defined as
\begin{equation}
    \vert W_n\rangle = \frac{1}{\sqrt{n}}(\vert100\ldots0\rangle+\vert010\ldots0\rangle+\ldots+\vert000\ldots1\rangle),
\end{equation}
this is actually the most trivial case of selecting 1 asset out of \emph{n}, $\binom{n}{1}$. The preparation procedure of \emph{W} state can be represented as $\vert W_n\rangle =U_n(NOT_1\otimes I_{n-1})\vert 0^n\rangle$, where $NOT_1\otimes I_{n-1}$ represents the \emph{NOT} gate operating on the highest qubit in the first time step, see figure \ref{fig:Wstate}, and \emph{$U_n$} is the \emph{n}-qubit unitary operator composed by the following ``Controlled-$R_y$, \emph{CNOT}'' (2C) blocks, hence is called CC ansatz, shown in the dashed box. The fact we found is that \emph{W} state can be regarded as being extracted from the column $2^{n-1}$ of the unitary \emph{$U_n$} as $\vert 1 0^{n-1}\rangle =NOT_1\otimes I_{n-1}\vert 0^n\rangle$.

Based on generalization and summarization, CC ansatz can be alternatively constructed in three steps when $k\leq n/2$. The first step is to deploy unitary $X_{\{k \operatorname{mod} 2\}}\otimes[(I\otimes X)^{\otimes\lfloor k/2 \rfloor}]\otimes I_{n-k} $ from top to bottom in the first time step, see figure \ref{fig:second ansatz} for the example of preparing $\vert D^5_2\rangle$. The subscript $\{k \operatorname{mod} 2\}$ means the first \emph{X} operates only when $k$ is odd. Then append a staircase composed of multiple 2C blocks for each \emph{X} gate except the one operating on the highest qubit (if exists). Each staircase starts from the corresponding \emph{X} gate and stops at the lowest qubit. At last append \emph{$U_n$} to complete the ansatz. The number of staircase layer is reduced to $\lfloor k/2\rfloor +1$ for preparing $\vert D^n_k\rangle$, whereas the prepared state with $k\in [3, n-3]$ contains non-target basis states. Same as the method used in extending to cases with larger $k$ in CCC ansatz, state $\vert D^n_k\rangle$ with $k>n/2$ in CC ansatz is prepared by $\vert D^n_{n-k}\rangle$. Each 2C block is also optimized to $2$ $CNOT$ gates, see appendix \ref{appen: optimal theory} for the two compile methods provided.

Although the state prepared by CC ansatz is not a precise Dicke state $\vert D^n_k\rangle$ and contains non-negligible non-targets when $k$ belongs to [$3$, $n-3$], the number of layers of staircase is reduced by almost half, making it more preferable for executing on NISQ computers. We also propose a symmetric space partition scheme that utilizes the reverse symmetry property of the basis states of the Dicke state to alleviate the impact of the extra basis states and preserve the target basis states as much as possible. See appendix \ref{appen: symmetric space partition scheme} for analysis and numerical illustrations.

The two ansatze both reach a linear complexity of two-qubit gate depth with a staircase structure, and $nk-3k^2/2$ and $n(k+1)/2-k^2/4$ parameters respectively, see  appendix \ref{appen: building block complexity analysis} for the evaluation of the complexity. And the comparison with the latest schemes for preparing Dicke states is presented in table \ref{tab:Dicke states schemes}. Concretely, to prepare the small scale Dicke state $\vert D^8_2\rangle$, the number of \emph{CNOT} gates required based on the 5 methods are 44, 27, 31, 22 and 25 respectively. Our ansatze achieve the minimum \emph{CNOT} counts with linear-nearest-neighbor coupling which is more friendly for the current NISQ computers. Compared to the previous circuits, the present ansatze are much more regular and simple. It is also worth noting that we can easily prepare a subspace of arbitrary $\vert D^n_k\rangle$ by combining fewer columns. In this manner, a much smaller and more accurate search space can be spanned when some prior knowledge of the solution is presented. Otherwise, we can still split the Dicke state into multiple subspaces and search them one by one to alleviate the issues of barren plateau (BP) and local minima \cite{Bittel2021,Tueysuez2022}. 

As can be seen, the ansatze are designed by reducing the parameters and quantum gates to minimum for improving their ability of executing on NISQ computers, while the expressibility is restricted, but the BP problem is also suppressed \cite{Holmes2022}. Intuitively for classical optimization problems in the NISQ era, the BP problem appears to be more important than expressibility. In addition, a local cost function can perform better than a global one \cite{Holmes2022, CerezoBP2021}. Here we use the conditional value-at-risk (CVaR) \cite{BarkoutsosCVaR2020} as the cost function. CVaR with a small confidence level $\alpha \in(0, 1]$ can be regarded as a local cost function. Additionally, the space spanned by the proposed ansatze always contain the ground states for all cases. Therefore, the negative impacts of restricted expressibility of the proposed ansatze can be greatly mitigated. We found numerically that deploying multiple layers of the staircase unitary $U_n$ composed of 3C blocks, which always prepares a precise Dicke state, really does not observably improve the performance for small confidence levels, see appendix \ref{appen: multi layers of Un for CCC ansatz} for the demonstrations. Therefore, in combinatorial optimization problems, it is reasonable for us to focus on the single layered $U_n$ with CVaR, which is more amenable to the current NISQ devices.

\section{Numerical simulations and experiments}
\label{sec:experi}

In this section, we perform the numerical simulations and hardware experiments to verify the effectiveness of our proposals.

\subsection{Numerical simulations}

We performed extensive noise-free numerical simulations by pyQPanda \cite{Dou2022}. Specifically, COBYLA algorithm is utilized. Simulation results show that we cannot expect to get the optimal solution in just one calculation when the problem has a little larger asset pool, especially for HE ansatz that spans an exponentially large search space. Therefore, it is more intuitive to statistically analyze and compare the overall performance of these ansatze. However, the variance is little larger due to the lower expressibility. In the following, we analyze numerically the time consumption and the probabilities of obtaining the optimal and feasible solutions for the linearly entangled HE ansatz and our proposed CCC and CC ansatze with respect to  different numbers of assets and budgets. From the function point of view, expectation $\langle \psi(\bm{\theta})\vert H \vert \psi(\bm{\theta})\rangle$ is a non-linear function of parameters $\bm{\theta}$. The performance of the algorithm should be crucially impacted by the initialization of the parameters $\bm{\theta}$ and the order of the assets in the present shallow ansatz scenario. In consequence, for obtaining a good solution with a high enough probability, the appropriate correspondence between them is taken into consideration. How the order of the assets extracted from the asset pools influences the probability of obtaining good solutions, rather than the widely investigated initialization of parameters, is also discussed.

\begin{figure}[ht]
    \begin{subfigure}{0.325\textwidth}
     \centering
     \includegraphics[width=\textwidth]{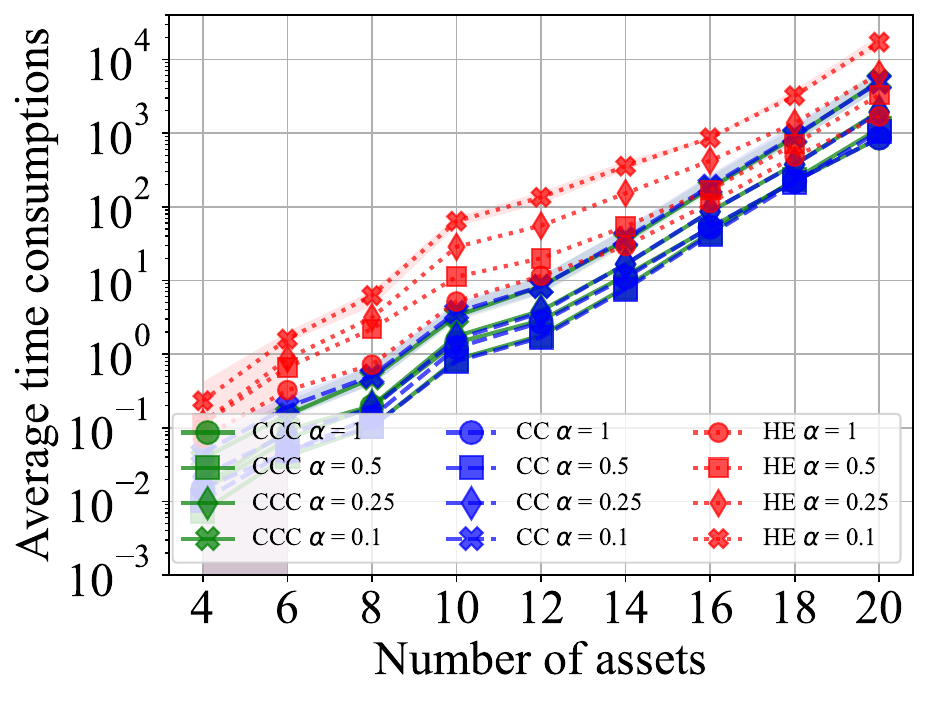}
     \caption{}
    \end{subfigure}
    \begin{subfigure}{0.325\textwidth}
     \centering
     \includegraphics[width=\textwidth]{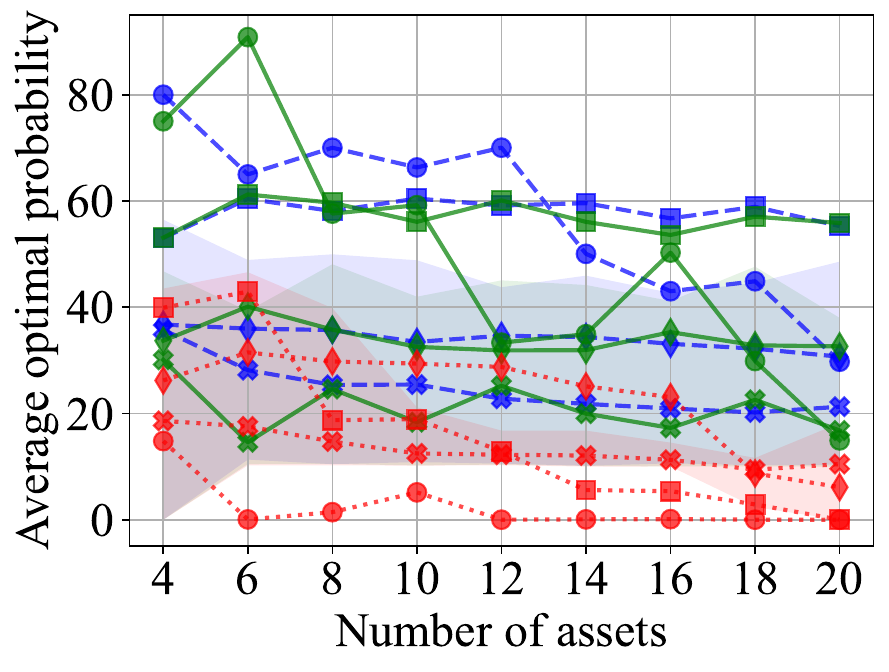}
     \caption{}
    \end{subfigure}
    \begin{subfigure}{0.325\textwidth}
     \centering
     \includegraphics[width=\textwidth]{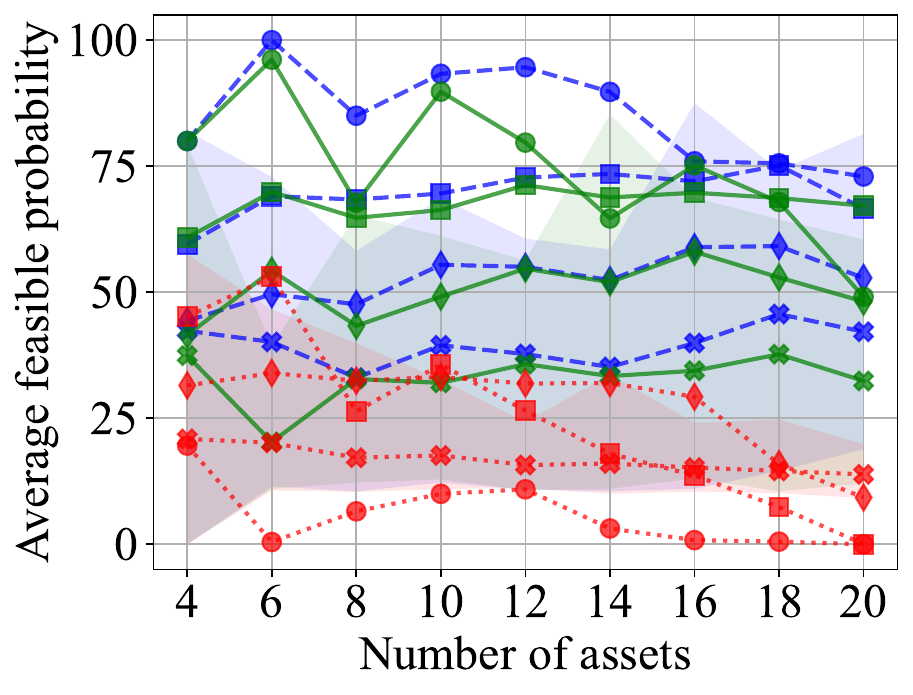}
     \caption{}
    \end{subfigure}
    \caption{\small The comparisons among the CCC, CC and HE ansatze, up to 20 assets: (a) the average time consumption (s), (b) the average probability of obtaining the optimal solution (\%), and (c) the average probability of obtaining the feasible solutions (\%). A feasible solution is accepted when its expectation is not more than 0.75 times that of the optimal solution. The maximum and minimum values at $\alpha$ = 0.1 are plotted as shadows. The others are not shown because they would override confusingly, disrupting the display of average information.\vspace{-1em}
}
    \label{fig:different assets}
\end{figure}

Figure \ref{fig:different assets} provides a sketch of the average time and probabilities of the three ansatze with the number of assets $n$ ranging in $[4, 20]$ and confidence level $\alpha\in\{1, 0.5, 0.25, 0.1\}$. The asset data is produced using $qiskit\_finance.data\_providers$ \cite{Qiskit}. The maximum number of iterations is set to 500. The number of budgets, $k$, is fixed at 2, and the risk level $q$ is set to 0.5. To illustrate the statistical behaviors of the time consumption and the probabilities, we initialize the parameters $\bm{\theta}$ with a random seed 1231 and the asset pool 20 times with random seeds starting from 1000. To guarantee shallow property, viable execution time and essential expressibility of the HE ansatz, the number of layers is set to be logarithmic in $n$. For consistency, the seeds used for the random initialization of the 20 asset portfolios are the same for each ansatz. To obtain results with the same precision, the number of samples of different confidence levels $\alpha$ should be $1/\alpha$ times of that of the cases where $\alpha = 1$. That is to say, the ability of always obtaining the optimal solution in most small $\alpha$ instances is achieved by appropriately increasing the sampling complexity. Due to the sampling taking much longer than classical post-processing, we roughly multipled the numerical simulation time by $1/\alpha$ to approximate the time consumption of each instance. Even so, under the same experimental conditions, the number of iterations trends to reduce with decreasing confidence level which  results in only a slight difference in the consumed time for different $\alpha$.

% \begin{multicols}{2}

As indicated in figure \ref{fig:different assets} (a), CCC and CC ansatze always take time less than the logarithmic depth HE ansatz. This benefits from the more targeted expressibility of our ansatze. The Hamming weight of the basis states prepared by our ansatze is exactly the budget $k$, while the HE ansatz outputs an exponential number of basis states that contain every possible budgets belonging to $[0, n]$. Then the search space spanned by our ansatze is much smaller, and the landscape is much more regular, which promises the high efficiency of our ansatze. In addition, to enhance the expressibility of the HE ansatz by deploying poly($n$) layers, the time consumption would increase rapidly.

Figure \ref{fig:different assets} (b) demonstrates the probabilities of obtaining the optimal solutions of the three ansatze. As the number of assets $n$ increases, the probabilities of obtaining the optimal solutions of CCC and CC ansatze are generally higher than that of the HE ansatz. And their performances improve as $\alpha$ decreases. For example, for the HE ansatz, $\alpha=0.5$ consistantly outperforms $\alpha=1$, $\alpha=0.25$ becomes superior to $\alpha=0.5$ when $n=8$, and $\alpha=0.1$ surpasses $\alpha=0.25$ starting from $n=18$. However, for the proposed ansatze, the upper bound is greatly improved. As shown in the figure, $\alpha=0.5$ does not surpass $\alpha=1$ until $n$ reaches approximately 10. Besides, the decreases in the cases corresponding to $\alpha=0.5,0.25,0.1$ are relatively slower, suggesting that the proposed ansatze exhibit high and stable performance for larger-scale cases, even with large $\alpha$. Conventionally, finding the optimal solution could be an extremely hard task that implies finding a feasible solution by consuming affordable resources is more realistic. As shown in figure \ref{fig:different assets} (c), compared to the probabilities of obtaining the optimal solution, CCC and CC ansatze can always find feasible solutions with much higher probabilities, especially for larger $n$. We argue that it is more practical to find a feasible solution for optimization problems using NISQ computers because the hardware noise and sampling error lead to the rapid convergence of the algorithm to a local minima that very likely corresponds to a feasible solution.

\begin{figure}[ht]
    \begin{subfigure}{0.325\textwidth}
     \centering
     \includegraphics[width=\textwidth]{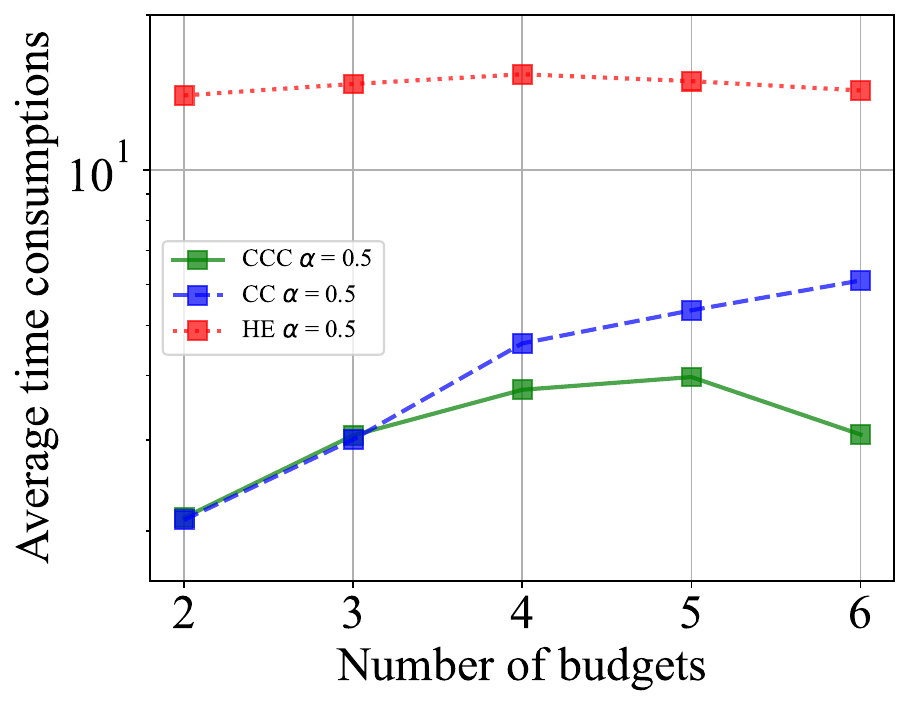}
     \caption{}
    \end{subfigure}
    \begin{subfigure}{0.325\textwidth}
     \centering
     \includegraphics[width=\textwidth]{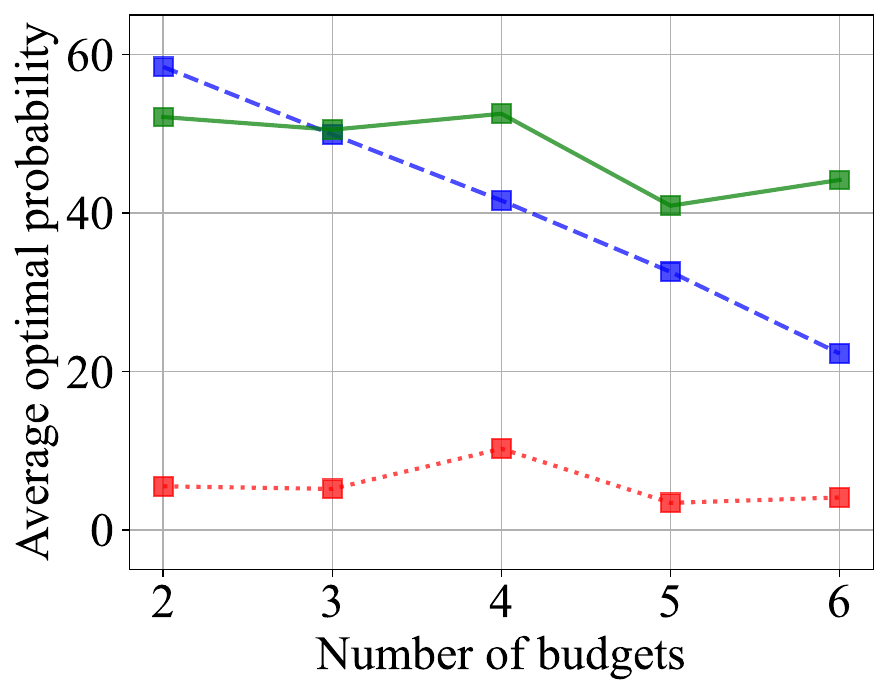}
     \caption{}
    \end{subfigure}
    \begin{subfigure}{0.325\textwidth}
     \centering
     \includegraphics[width=\textwidth]{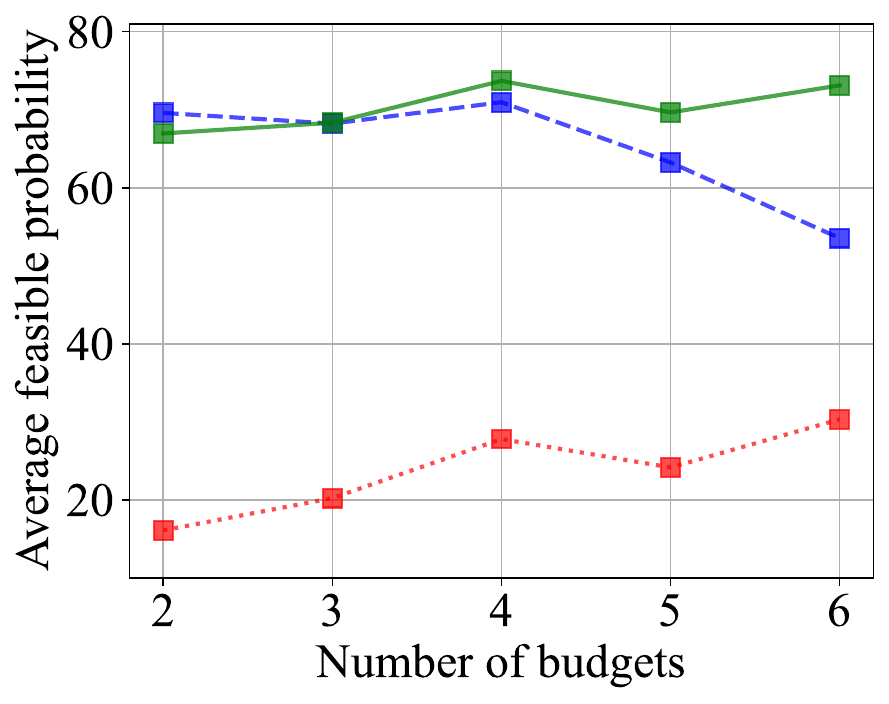}
     \caption{}
    \end{subfigure}
    \caption{\small The comparisons of different number of budgets: (a) the average time consumption (s), (b) the average probability of obtaining the optimal solution (\%), and (c) the average probability of obtaining the feasible solutions (\%).\vspace{-1.6em}}
    \label{fig:different budgets}
\end{figure}
\begin{figure}[ht]
    \begin{subfigure}{0.325\textwidth}
     \centering
     \includegraphics[width=\textwidth]{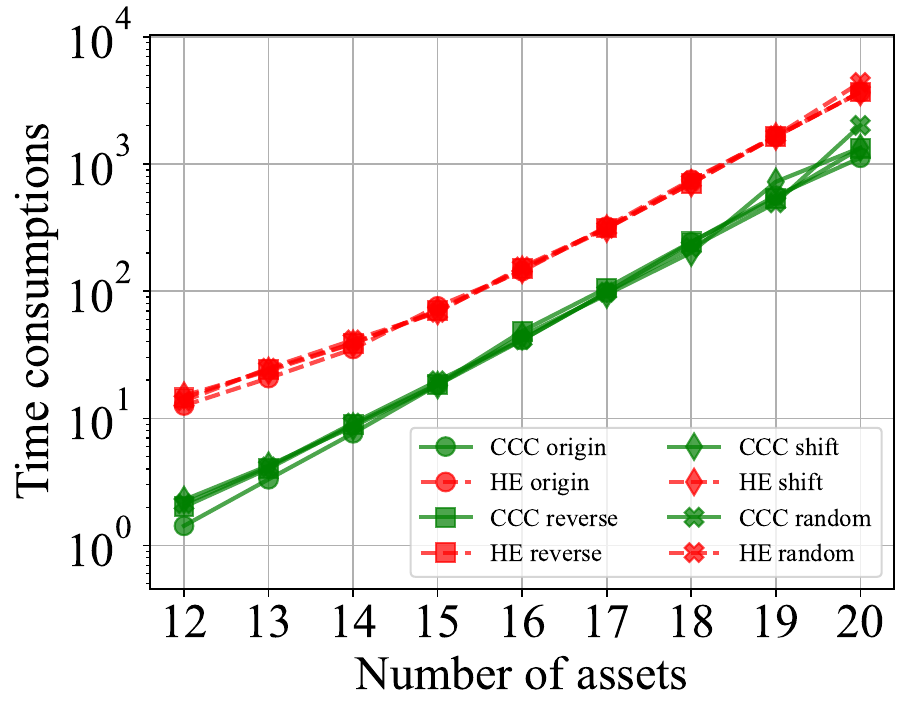}
     \caption{}
    \end{subfigure}
    \begin{subfigure}{0.325\textwidth}
     \centering
     \includegraphics[width=\textwidth]{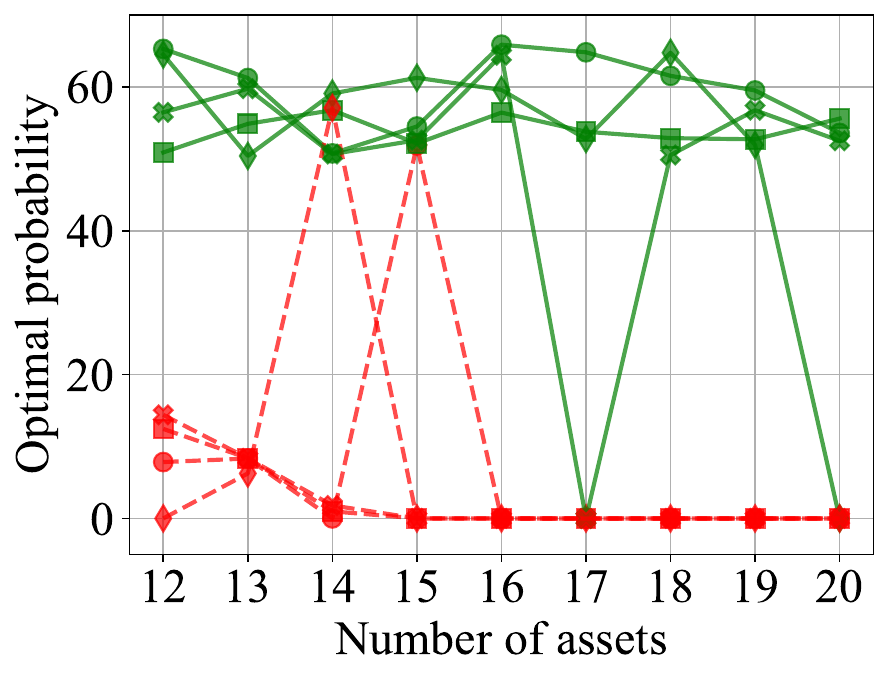}
     \caption{}
    \end{subfigure}
    \begin{subfigure}{0.325\textwidth}
     \centering
     \includegraphics[width=\textwidth]{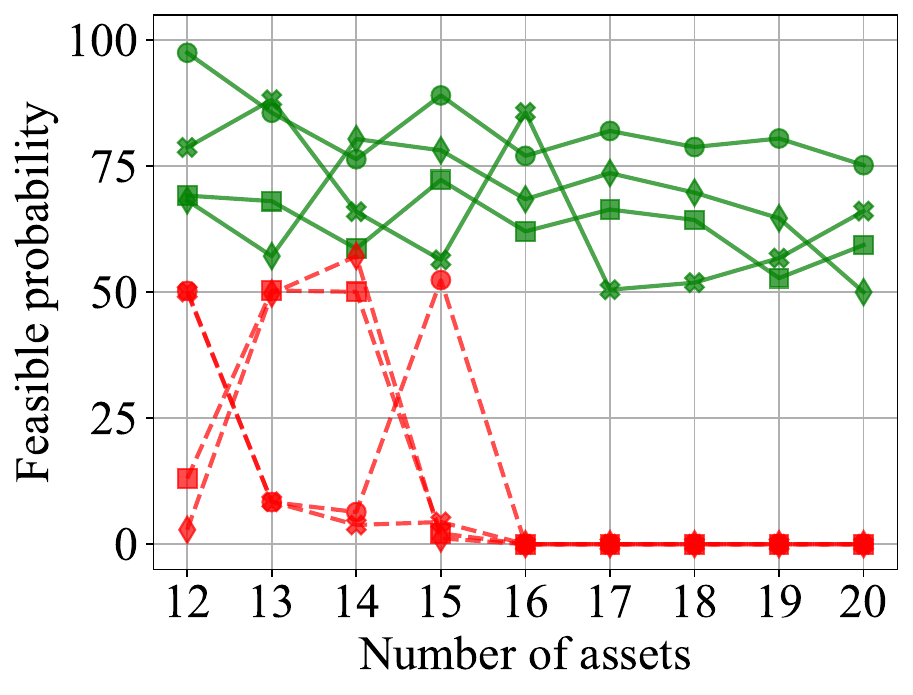}
     \caption{}
    \end{subfigure}
    \caption{\small The comparisons of the four kinds of asset orders: (a) the time consumption (s), (b) the probability of obtaining the optimal solution (\%), and (c) the probability of obtaining the feasible solutions (\%). CC ansatz is not shown due to its similar performance to CCC ansatz.\vspace{-1em}}
    \label{fig:different orders}
\end{figure}

In the second simulation, we fix confidence level $\alpha$ at 0.5, the total number of assets at 12, and test the performance of these ansatze when the budgets range from 2 to 6. As shown in figure \ref{fig:different budgets}, the performance of the proposed ansatze is always much better than the HE ansatz. In figure \ref{fig:different budgets} (a), the time consumed by the HE ansatz remains almost the same because the ansatz is the same one across different budgets, while the time consumed by the proposed ansatze gradually increases due to the expansion of the search space. In figure \ref{fig:different budgets} (b), the probability of obtaining the optimal solution of CC ansatz decreases much faster than for CCC ansatz due to the introduction of extra non-target states. Compared to the probability of obtaining the optimal solution, it is much easier for us to obtain a feasible solution, as depicted in figure \ref{fig:different budgets} (c).

At last, the effects of four different orders of the assets, namely the original order, reversed order, shift order and random order, rather than altering the initialization of parameters $\bm{\theta}$, are compared by setting the value of budget $k$ and confidence level $\alpha$ to 2 and 0.5 respectively. In this scenario, the parameters and the assets are initialized only once to indicate the complementary behavior of the four different orders. The results are depicted in figure \ref{fig:different orders}, with a time consumption similar to that of the previous two simulations. For CCC ansatz, there always exits at least one order that samples the optimal solution with a high enough probability. However, for the cases with more than 16 assets, the HE ansatz cannot even sample a feasible solution. In such cases, we may need a much smaller confidence level to ``amplify'' the probability.

\subsection{Experiments on NISQ computer}

The common fact is that, to achieve potential quantum advantage, the problem scale needs to be large enough. That is to say, the hardware should be able to prepare a high-precision superposition state containing an exponential number of basis states with respect to a large number of qubits. However, the current NISQ computers cannot easily process a large number of qubits, while retaining an exponential number of basis states simultaneously.

To execute the ansatze efficiently on the current NISQ computers, the quantum/classical hybrid distributed computing (HDC) scheme is proposed. The feasibility of HDC scheme stems from the fact that we can find a combination of a small number of columns that encompass all target basis states. Then we split classically the ansatz into several subcircuits, each of which extracts the target from one or a few of these columns. In other words, each subcircuit is synthesized using only one of these columns, or a superposition of a few of these columns, if possible, followed by $U_n$. At this time, each subcircuit with relatively sparse entanglement spans a smaller subspace that is more applicable for NISQ computers. The qubit reuse technique \cite{Hua2022} can be applied to execute these subcircuits directly, especially the subcircuits with very few columns, see figure \ref{fig:single layer of staircase} in  appendix \ref{appen: XY reduction}. In this paper, we focus on the more studied circuit cutting technique that can mitigate the impact of errors more flexibly in the width dimension of NISQ devices \cite{Peng2020,Wang2021}. By careful design, the sparse staircase structure of the subcircuits can offer us superior properties like minimal number of cut qubits and correspondingly minimal number of observables of the ``measure-and-prepare'' channels for utilizing the circuit cutting technique. To be specific, the total number of cut qubits for partitioning each subcircuit into $p+1$ fragments is just $p$ and the ``measure-and-prepare'' channel only has the observables $\vert 0\rangle \langle 0\vert$ and $\vert 1\rangle \langle 1\vert$ when a single qubit is cut between the adjacent two fragments, see appendix \ref{appen: XY reduction} for the detailed proof. Then the sampling complexity of each subcircuit for evaluating the $\operatorname{CVaR}_{\alpha}$ is $O(1/(\alpha\epsilon^2))$ with $\alpha \in(0, 1]$ the confidence level and $\epsilon$ the accuracy. This is a complexity that has no concern with the total number of cut qubits $p$. Hence the ansatz can be partitioned into more low-bit fragments for obtaining the low-probability basis states more easily when the subcircuit is suitable. This is because each piece of the basis states can be obtained with higher probabilities in the corresponding fragment, leading to more accurate results.

The reduction of the ``measure and preprare'' channels leads to the sequential sampling of the fragments. The sampling procedure 
is as follows: (1) The first step is to measure the first fragment with the number of shots, say $N$, satisfying the sampling requirement, Hoeffding's inequality \cite{Peng2020}, on matrix basis $I$. The $N$ measured strings are stored one by one according to the order in which they are measured and the number of the values 0 and 1 of the cut qubit are counted as $N_0$ and $N_1$. (2) At the second step, the second fragment with input 0 and 1 are measured $N_0$ and $N_1$ times respectively. Based on the values of the cut qubit of the first $N$ strings, the measured strings of the second fragment are appended to the first $N$ strings one by one in accordance with their measured order. $N_0$ and $N_1$ are updated to be the number of the values 0 and 1 of the cut qubit of the second fragment. (3) Repeat step (2) until all fragments are appended. For instance, the measured strings of the first fragment are $01\bm{0}$, $01\bm{1}$, $11\bm{0}$, $10\bm{1}$ with the last qubit being the cut qubit. Then the second fragment with inputs 0 and 1 is executed 2 times respectively. Assume the measured strings are $11_0$, $01_0$, and $11_1$, $00_1$ based on input 0 and 1 (represented as subscript) respectively, then the order corresponding to the first strings is $11_0$, $11_1$, $01_0$, $00_1$. So the combined basis states are 0111, 0111, 1101, and 1000. For $p+1$ fragments, the number of times we submit a new circuit to the hardware decreases to $2p+1$. Here we propose to simultaneously sample each fragment, which is also advantageous for applying quantum error mitigation (QEM) \cite{Cai2022,Bonet-Monroig2018,Nation2021,Temme2017}. In this case, all fragments are shot $N$ times simultaneously. Then we perform the sequential sampling of the measured results in a classical manner. 

The states prepared by problem-inspired ansatze are sparse ones, so the commonly laudable noise-resistance feature is no longer as effective. Here by combining the symmetry property of the states prepared by each fragment with measurement error mitigation method \cite{Cai2022,Nation2021}, we greatly improve the accuracy of the execution on NISQ hardwares. Specifically, we found that each fragment can only output states with the same Hamming weight, which provides us with a more accurate assignment matrix $M$ by setting the columns with other Hamming weights to be a standard basis with the element ``1'' on the main diagonal. Then after evaluating $\Vec{p}_{migit} = M^{-1}\Vec{p}_{noisy}$ where $\Vec{p}_{noisy}$ is the vector of the measured probability and $\Vec{p}_{migit}$ is the mitigated probability, we normalize the probabilities of the states with the correct Hamming weight in $\Vec{p}_{migit}$ to be the final output. Our proposals mitigate the effects of noise perturbation and, as a result, improve the stability and convergence speed of the optimization process. Our ansatze are insensitive to phase flip errors in the scenarios involving a single layer of staircase structure, which can be relaxed to suppress the more critical bit flip errors.

\begin{figure}[ht]
    \centering
    \includegraphics[width=0.325\textwidth]{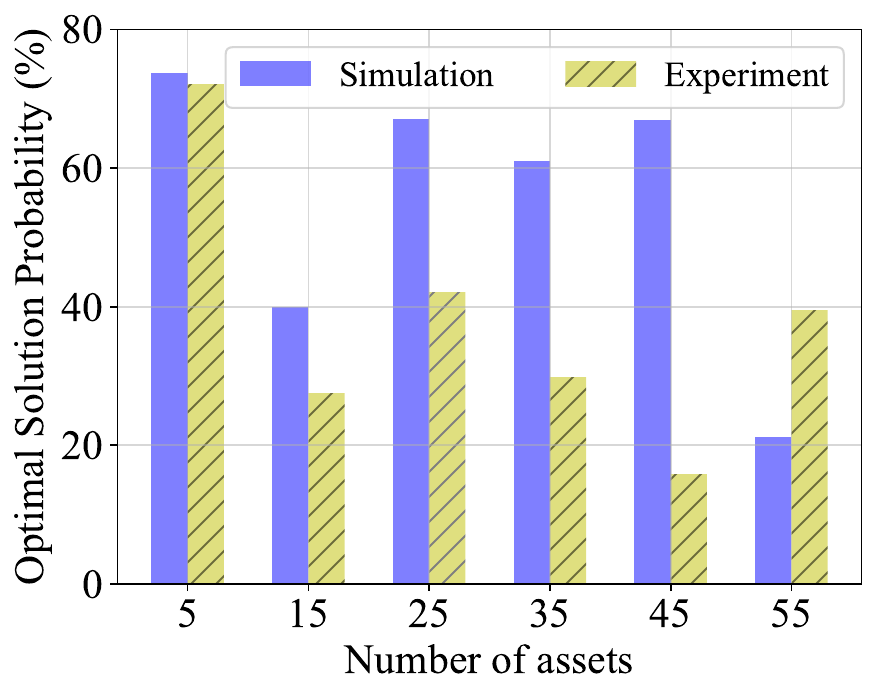}
    \caption{The probability of obtaining the optimal solution in the first experiment, numerical versus hardware. Appendix \ref{appen: experimental setup} shows the simulation and experiment setups, and qubit topology for both experiments. In this experiment, qubits 45, 46, 52 are used. It should be clarified that we did not perform experiments with qubits larger than 55.}
    \label{fig:theory_experi_comp}
\end{figure}

In addition, the change of each $\theta$ only influences the corresponding fragment, and all other fragments remain unchanged, so there is no need to execute them repeatedly. This fact can be utilized to reduce the number of submissions when classical sampling is allowed, and the introduced bias can improve the stability of the optimization process. Hence the trainability can be improved as well. We call it ``fragment reuse technique''.

The first experiment executes the $k=3$ case of CCC ansatz with one column corresponding to one subcircuit on the NISQ computer to demonstrate the feasibility of the proposed algorithms; CC ansatz has similar results. For the specific hybrid distributed design schemes and complexity analysis of $k=2$ and $3$ cases, see appendix \ref{appen: Experiment schemes and complexity}. Each ansatz is partitioned into multiple 3-qubit fragments, and the asset pool is generated by using random seed 1000. To improve the convergence speed, the confidence level $\alpha$ is initially set to a small value, as shown in table \ref{tab:Setup of experiments} of appendix \ref{appen: experimental setup}, and is then increased to 1.0 gradually \cite{Kolotouros2022}. We also formalize the uniform initialization of the states for all subcircuits. %, see the code attached for details

In this scenario, SLSQP is employed, rather than COBYLA, because this experiment needs to handle many small-size fragments, and SLSQP can take better advantage of fragment reuse technique to improve the stability of the optimization process (cost gradient). In addition, there are only 3 ``1'' elements in each basis state, and their distribution is relatively dispersed. The gradient of the cost function is easier to find. The experiments achieve the successful executions of the circuits with up to 55 qubits. Figure \ref{fig:theory_experi_comp} shows the probability comparison between numerical simulation and hardware experiment. The optimal solutions in both cases are obtained with high enough probabilities. Furthermore, hardware experiments converge much faster under the same configuration, which possibly stems from the perturbation effect of the hardware noise.

\begin{figure}[h]
    \centering
    \includegraphics[width=0.39\textwidth]{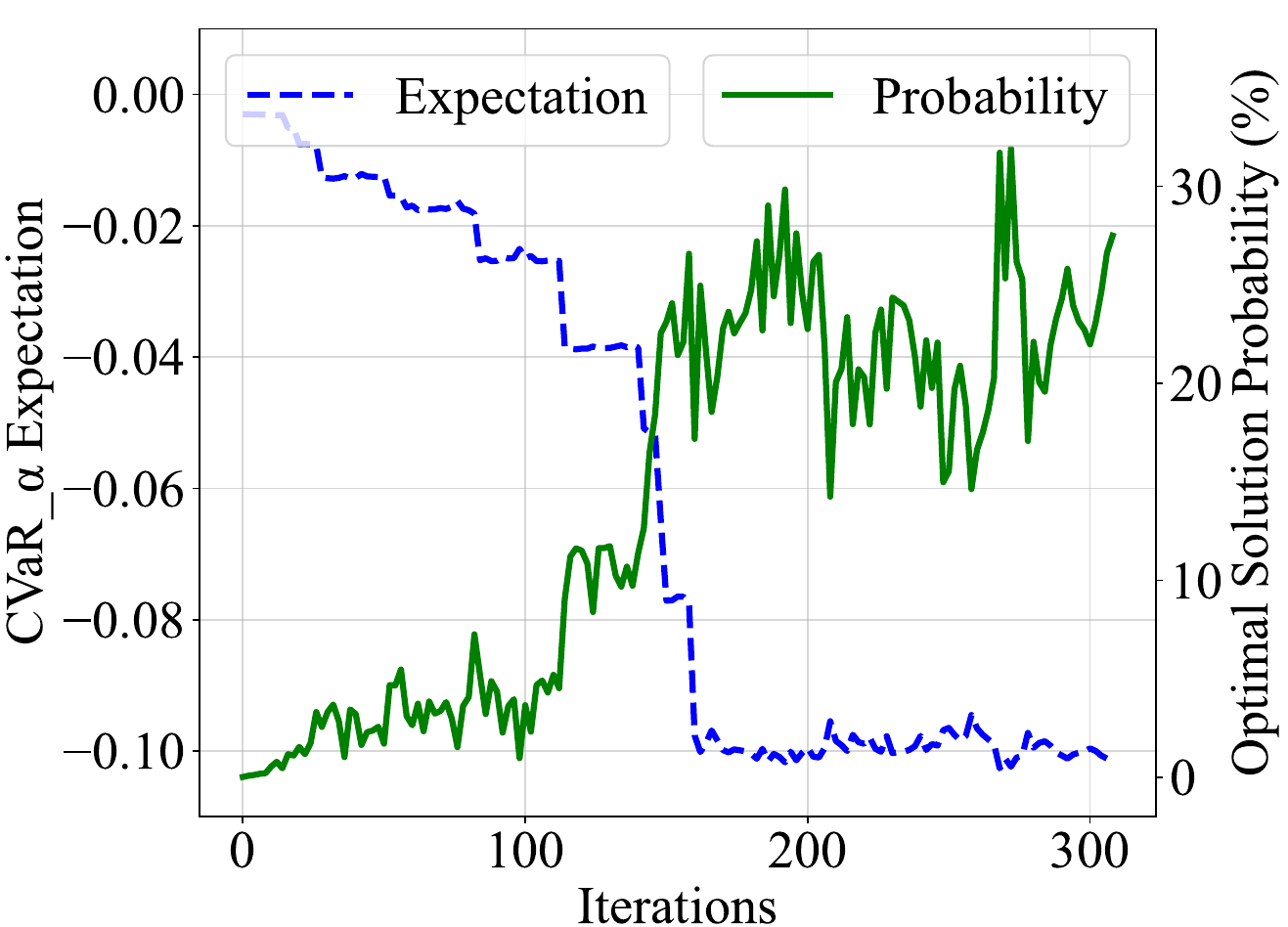}
    \caption{The convergence curve for case $\vert D^{12}_{6} \rangle$. The blue dashed line shows the change of the CVaR expectation, and the green solid line represents the probability of obtaining the optimal solution.}
    \label{fig: c12,6_experi_prob and expect curve}
\end{figure}

The second experiment searches for the optimal solution of case $\vert D^{12}_{6} \rangle$. 6 qubits are employed in each fragment. The details of the experimental scheme are presented at the end of appendix \ref{appen: Experiment schemes and complexity}. At this time, the restricted expressibility and complex structure of our ansatz severely impede its uniform initialization. To remedy this issue, we coarsely fit the parameters $\bm{\theta}$ that can produce a relatively uniform output. The coarse fitting is performed by minimizing $\sum_i(p_i^{'}-p_0)^2$ where $p_i^{'}$ corresponds to the probabilities of the sampled states and $p_0$ is the uniform probability. Additionally the parameters $\bm{\theta}$ are also initialized randomly in the interval $[\pi/2, 3\pi/4]$.

Extensive numerical simulations showed that, in the most complicated scenario, COBYLA outperforms SLSQP. As shown in figure \ref{fig:different budgets} (b), when using COBYLA and setting $\alpha$ to 0.5, the average probability of obtaining the optimal solution of $\vert D^{12}_{6} \rangle$ is about 45\% (approaching the soft cap of 50\% corresponding to $\alpha=0.5$), while with the same configuration, SLSQP can only achieve 18\%. This results from the BP and local minima problems caused by the considerable overlap of the basis states. Specifically, the asset pool generated by the random seed 1000 also prevents SLSQP from converging to the optimal solution. The present experimental scheme can only slightly improve the situation. In such a difficult instance, we attempted to use COBYLA. In the primary stage of COBYLA, only one or two parameters are changed in each iteration. So, we choose to repeatedly run the primary stage multiple times. The parameters obtained from the previous run are used as the initial parameters of the next run. In consequence, the fragment reuse technique can make the optimization process more stable. As shown in figure \ref{fig: c12,6_experi_prob and expect curve}, the optimization process exhibits fluctuating convergence to the optimal solution. The fluctuation can mainly be attributed to the learning rate adjustment made at the beginning of each run and the unstable output results of the hardware. Additionally, the optimization process of COBYLA itself can also contribute to mild fluctuations.

\section{Conclusion and Discussion}
\label{sec:disc&concl}
In this paper, we have achieved the common portfolio optimizations using a superconducting computer for up to 55 qubits (assets) in the case of $\vert D^{n}_{3} \rangle$ and 12 qubits in the case of $\vert D^{n}_{n/2} \rangle$. This comes from trading expressibility for trainability by proposing two compact Dicke state ansatze and utilizing the CVaR cost function. Additionally, the proposed HDC scheme with simultaneous sampling, problem-specific measurement error mitigation, fragment reuse technique, is utilized. The proposed ansatze achieve the current lowest complexity in terms of circuit depth, two-qubit gates, and parameters, compared to previous methods. And the unified framework for preparing arbitrary Dicke states with a staircase structure is well-suited for distributed quantum computing. Our proposals greatly improve the execution precision in NISQ devices with little overhead.

It should be noted that we just applied the simplest quantum error mitigation method, i.e. quantum measurement error mitigation. Other superior methods, such as zero-noise extrapolation and probabilistic error cancellation \cite{Cai2022}, should be able to achieve significantly better results. Hence, as the performance of NISQ computers improves, the methods we proposed have the potential to significantly accelerate certain practical applications of variational quantum algorithms \cite{Cerezo2021}, such as in the fields of chemistry, biology, and machine learning, in the near future, not limited to classical optimization problems.

The experimental results show that partitioning the search space into multiple small-enough subspaces can alleviate the barren plateau problem. Nevertheless, to obtain potential quantum advantages, ergodically searching these subspaces may not be efficient in some cases. Additionally, the error-prone issue of NISQ computers limits their capability further. In consequence, choosing the appropriate columns or partial elements of these columns by consuming few resources to span a much accurate search subspace becomes critical. However, it maybe cumbersome due to its dependence on the correlation among the assets and subspaces.

\section*{Data availability statement}
The data that support the findings of this study are available upon reasonable request from the authors.

\section*{Acknowledgement}
This work was supported by the National Natural Science Foundation of China (Grant No. 12034018), by the Innovation Program for Quantum Science and Technology (Grant No. 2021ZD0302300).

\newpage
\appendix

\counterwithin{figure}{section}
\counterwithin{table}{section}
\counterwithin{equation}{section}
\begin{appendices}

\section{Theoretical Analysis}
\label{appen: theoretical support}

\renewcommand{\thefigure}{A\arabic{figure}}
\renewcommand{\thetable}{A\arabic{table}}
\renewcommand{\theequation}{A\arabic{equation}}
In this appendix, we provide theoretical insights to some extent for why the proposed ansatze are capable of preparing arbitrary Dicke states efficiently with fairly low complexity. In the main text we show that CC ansatz is inspired directly by the structure of the circuit (figure \ref{fig:Wstate}) for preparing \emph{W} state. That is to say, the case of $\vert D^n_1\rangle$ in Dicke states is prepared by extracting the information stored in column $2^{n-1}$ of unitary $U_n$. In the following, we take $\vert D^5_k\rangle$ as the warm-up example to generalize the intuition for preparing arbitrary Dicke states. This idea is also likely to provide insights into why matrix product states and multiple layers of hardware-efficient (HE) ansatz can efficiently approximate quantum states.

\begin{figure}[ht]
    \centering
    \begin{subfigure}{0.4\textwidth}
     \centering
     \includegraphics[width=\textwidth]{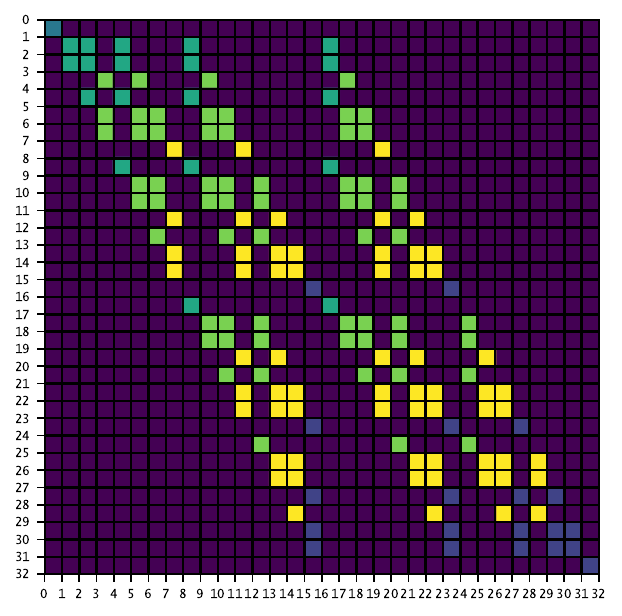}
     \caption{}
    \end{subfigure}
    \begin{subfigure}{0.4\textwidth}
     \centering
     \includegraphics[width=\textwidth]{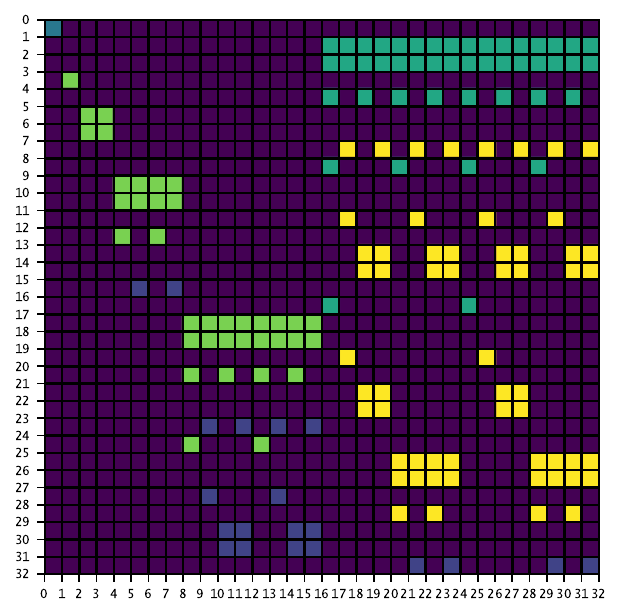}
     \caption{}
    \end{subfigure}
    \caption{The matrices for $U_5$ based on (a) 3C and (b) 2C blocks.}
    \label{fig:Unmatrices}
\end{figure}

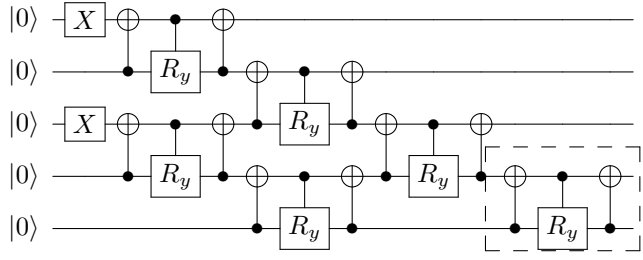
\begin{figure}[ht]
    \centerline{
    \Qcircuit @C = 0.45em @R = 0.4em  @!R{
     &\lstick{\vert0\rangle} &\gate{X}  &\targ     &\ctrl{1}   &\targ      &\qw       &\qw        &\qw        &\qw       &\qw        &\qw       &\qw       &\qw        &\qw       &\qw \\\
     &\lstick{\vert0\rangle} &\qw       &\ctrl{-1} &\gate{R_y} &\ctrl{-1}  &\targ     &\ctrl{1}   &\targ      &\qw       &\qw        &\qw       &\qw       &\qw        &\qw       &\qw \\\
     &\lstick{\vert0\rangle} &\gate{X}  &\targ     &\ctrl{1}   &\targ      &\ctrl{-1} &\gate{R_y} &\ctrl{-1}  &\targ     &\ctrl{1}   &\targ     &\qw       &\qw        &\qw       &\qw \\\
     &\lstick{\vert0\rangle} &\qw       &\ctrl{-1} &\gate{R_y} &\ctrl{-1}  &\targ     &\ctrl{1}   &\targ      &\ctrl{-1} &\gate{R_y} &\ctrl{-1} &\targ     &\ctrl{1}   &\targ     &\qw \\\
     &\lstick{\vert0\rangle} &\qw       &\qw       &\qw        &\qw        &\ctrl{-1} &\gate{R_y} &\ctrl{-1}  &\qw       &\qw        &\qw       &\ctrl{-1} &\gate{R_y} &\ctrl{-1} &\qw \\\ \gategroup{4}{13}{5}{15}{1.4em}{--}
    }}
    \caption{Illustration of the unsimplified version for preparing $\vert D^5_2\rangle$. The phase gates can be eliminated when 3C blocks are transpiled into the optimal circuits, see appendix \ref{appen: optimal theory} for the optimal circuits.}
    \label{fig:first ansatz no reduction}
\end{figure}

For CCC ansatz, the matrix of $U_5$ is shown in figure \ref{fig:Unmatrices} (a), where zero elements are indicated by the purple squares, and non-zero elements are represented by other colored squares. As can be seen, the Hamming weight of the elements in each column is unique. Hence we can prepare arbitrary $\vert D^n_k\rangle$ precisely. Specifically, (I) the columns $8=\bm{01}000_2$ and $16=\bm{10}000_2$ indicated by cyan can both prepare $\vert D^5_1\rangle$. This is because the shift of the $X$ gate between the highest and second highest qubits in the upper left region in figure \ref{fig:first ansatz no reduction} only induces a bit flip for the following $R_y$ rotation. (II) For the green case of $k=2$, there are multiple choices that are the set of columns $9=\bm{01}001_2, 12=\bm{01}100_2$ or set of columns $17=\bm{10}001_2, 20=\bm{10}100_2$. For a quantum device with richer qubit connectivity beyond just linear-nearest-neighbour coupling, each set of columns could be prepared directly without introducing SWAP operations. Without loss of generality here we add $18=\bm{10}010_2$ to the second set and form an exponential sequence composed of $1, 2, 4$ with the highest qubit fixed to be 1 to complete figure \ref{fig:first ansatz no reduction}. It can be found that the sequence of $1, 2, 4$ forms a $\vert D^3_1\rangle$ state. (III) The $k=3$ case represented by yellow squares can be summarized in the same manner by inserting column $21=\bm{10}101_2$ to the set composed of columns $19=\bm{10}011_2$ and $22=\bm{10}110_2$. The little difference is that the 3 low bits form a $\vert D^3_2\rangle$ state that is also a $k=2$ case with $n=3$ now. To sum up, for preparing target state $\vert D^n_k\rangle$, we need to prepare state $\vert D^{n-2}_{k-1}\rangle$ on the low $n-2$ bits and $\vert 10\rangle$ on the highest 2 bits first; then for preparing intermediate state $\vert D^{n-2}_{k-1}\rangle$, we need to prepare $\vert D^{n-4}_{k-2}\rangle$ on the low $n-4$ bits and $\vert 10\rangle$ on the highest 2 bits of the $(n-4)$ bits; \ldots, in a recursive way. The process can be formalized as
\begin{equation}
\begin{split}
\vert D^n_k\rangle= \ &U_n\vert 10\rangle\vert D^{n-2}_{k-1}\rangle, \\
\vert D^{n-2}_{k-1}\rangle= \ &U_{n-2}\vert 10\rangle\vert D^{n-4}_{k-2}\rangle,\\
&\quad\vdots\\
\vert D^{n-2(i+1)}_{k-(i+1)}\rangle= \ &U_{n-2(i+1)}\vert 10\rangle\vert D^{n-2i}_{k-i}\rangle,
\end{split}
\end{equation}
until one of the conditions $n-2i=k-i$ and $k-i=0$ is reached.

This circuit differs slightly from figure \ref{fig:first ansatz} because the last 3C block in the dashed box (see figure \ref{fig:first ansatz no reduction}) is eliminated. Through a similar inspection, we establish the commonly constructive framework as presented in the main text. The reason for further eliminating the 3C blocks stems from the fact that multiple layers introduce additional symmetries that alter only the amplitudes, not the completeness, of the basis states. As analyzed above, we have diverse choices (different sets of equivalent columns) for constructing the same Dicke states with various circuit structures. Our hybrid distributed ansatze are determined in this manner, see section \ref{sec:experi} and appendix \ref{appen: Experiment schemes and complexity}.

The discussion of CC ansatz is in a similar way. The matrix of $U_5$ is shown in figure \ref{fig:Unmatrices} (b). (I) As shown in the upper right of the figure, the cyan squares indicate the vectors with Hamming weight 1, i.e. the case of $k=1$. There are 16 columns containing $\binom{5}{1}$, but just columns $16=2^4$ and 24 contain all of the five target vectors without redundancy. This corresponds to the case of preparing $\vert W_5\rangle$ state, see figure \ref{fig:Wstate}. For the sake of uniformity and simplicity, column $16$ has been chosen. (II) The row indices flagged by the green squares in the left side are the vectors with $k=2$. The target basis states are dispersed in columns $\bm{1}$, ($\bm{2}$, 3), ($\bm{4}$, 6), ($\bm{8}$, 12) where the columns in the same parentheses contain the same target basis states. Obviously the exponential sequence composed of columns 1, 2, 4, 8 that is the $\vert W_4\rangle$ state perfectly constructs state $\vert D^5_2\rangle$, see figure \ref{fig:second ansatz}. That is to say, the linear combination required for extracting $\vert D^5_2\rangle$ is $\vert 0\rangle \vert W_4\rangle$. (III) The yellow squares in the right side denote the $k=3$ case where columns $17=16+\bm{1}, 18=16+\bm{2}, 20=16+\bm{4}$ form the exponential sequence 1, 2, 4 by fixing the value of the highest qubit to 1. Similarly the exponential sequence 1, 2, 4 form the state $\vert W_3\rangle$ on the low $3$ bits. The combined superposition state is $\vert 10\rangle \vert W_3\rangle$ now. Therefore CC ansatz is constructed based on this generalized pattern, i.e. $\vert D^n_k\rangle=U_nS^n_k\vert 0^n\rangle$ where $S^n_k$ is the linear combination of columns for extracting $\vert D^n_k\rangle$ from $U_n$. $S^n_k$ can be formalized for odd $k$ as

\begin{equation}
\begin{split}
S^n_1= \ &NOT_1\otimes I_{n-1}, \\
S^n_3= \ &(I_2\otimes U_{n-2})[(NOT_1\otimes I_1) \otimes(NOT_1\otimes I_{n-3})], \\
S^n_5= \ &(I_2\otimes U_{n-4})(I_4\otimes U_{n-4})[(NOT_1\otimes I_1) \otimes(NOT_1\otimes I_1)\otimes(NOT_1\otimes I_{n-5})], \\
&\quad \ldots \\
S^n_k= \ &[\prod^{\frac{k-3}{2}}_{i=0} (I_{(k-1)-2i}\otimes U_{n-(k-1)})] \{[\otimes^{\frac{k-3}{2}}_{i=0} (NOT_1\otimes I_1)]\otimes (NOT_1\otimes I_{n-k})\},
\end{split}
\end{equation}
and for even $k$ as

\begin{equation}
\begin{split}
S^n_0= \ &I_n, \\
S^n_2= \ &(I_1\otimes U_{n-1})(I_1\otimes NOT_1\otimes I_{n-2}), \\
S^n_4= \ &(I_1\otimes U_{n-3})(I_3\otimes U_{n-3}) [I_1\otimes NOT_1\otimes I_1\otimes NOT_1\otimes I_{n-4}], \\
&\quad \ldots \\
S^n_k= \ &[\prod^{\frac{k-2}{2}}_{i=0} (I_{(k-1)-2i}\otimes U_{n-(k-1)})] \{[\otimes^{\frac{k-2}{2}}_{i=0} (I_1\otimes NOT_1)]\otimes I_{n-k}\}.
\end{split}
\end{equation}

\noindent However this state is not the exact $\vert D^n_k\rangle$ in the cases of $2<k<n-2$ because a few extra basis states with Hamming weight not equalling $k$ are introduced. For solving practical portfolio optimization with small $k$ on NISQ computers, this is not the key problem. The strategies for mitigating this problem are presented in the main text and appendix \ref{appen: symmetric space partition scheme}. The critical is that CC ansatz has almost half the number of layers compared to CCC ansatz, which appears to be a beneficial characteristic. Furthermore, it provides us with a flexible way to regularly select the targets from a subspace that contains cases with multiple Hamming weights.

\section{Optimal Circuits for the Building Blocks}
\label{appen: optimal theory}
\renewcommand{\thefigure}{B\arabic{figure}}
\renewcommand{\thetable}{B\arabic{table}}
\renewcommand{\theequation}{B\arabic{equation}}
In the optimal two-qubit circuit theory \cite{Vatan2004}, a two-qubit quantum gate $U\in \textbf{SO}(4)$ can be constructed by $2$ \emph{CNOT} gates and at most 12 elementary one-qubit gates in the magic basis
\begin{equation*}
\mathcal{M}=\frac{1}{\sqrt{2}}
\begin{pmatrix}
1 &0  &i  &0 \\
0 &i  &0  &-1 \\
0 &i  &0  &1\\
1 &0  &-i &0
\end{pmatrix},
\end{equation*}
with its circuit

{\centerline{
\Qcircuit @C = 1em @R = 0.5em @!R{
&\gate{S} &\gate{H} &\ctrl{1} &\qw \\
&\gate{S} &\qw      &\targ    &\qw 
}}}
\vspace{1em}

However, the optimal circuit for the 2C block contains 3 $CNOT$ gates because the 2C block belongs to $\textbf{O}(4)$ with det(2C)$=-1$ \cite{Vatan2004}. We set the determinant of 2C block to 1 by replacing the $CNOT$ gate with $Controlled-R_y(\pi)$.

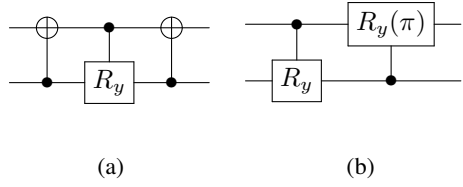
\begin{figure}[ht]
    \centering
    \begin{subfigure}{0.15\textwidth}
     %\centering
     \Qcircuit @C = 1em @R = 0.5em  @!R{
&\targ     &\ctrl{1}   &\targ     &\qw \\
&\ctrl{-1} &\gate{R_y} &\ctrl{-1} &\qw \\
\quad
     }
     \caption{}
    \end{subfigure}
    \begin{subfigure}{0.21\textwidth}
     \centering
     \Qcircuit @C = 1em @R = 0.5em  @!R{
&\quad &\ctrl{1}   &\gate{R_y(\pi)} &\qw \\
&\quad &\gate{R_y} &\ctrl{-1}       &\qw \\
\quad
     }
     \caption{}
    \end{subfigure}
    \caption{The building blocks of (a) 3C and (b) 2C.}
    \label{fig:3C and 2C}
\end{figure}

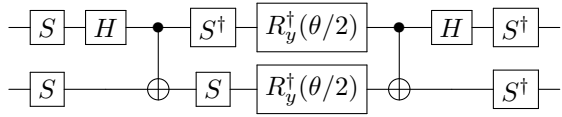
\begin{figure}[h]
    \centerline{
     \Qcircuit @C = 0.8em @R = 0.5em @!R{
     &\gate{S} &\gate{H} &\ctrl{1} &\gate{S^{\dagger}} &\gate{R_y^{\dagger}(\theta/2)} &\ctrl{1} &\gate{H} &\gate{S^{\dagger}} &\qw\\\
     &\gate{S} &\qw      &\targ    &\gate{S}           &\gate{R_y^{\dagger}(\theta/2)} &\targ    &\qw      &\gate{S^{\dagger}} &\qw 
     }}
    \caption{The optimal circuit for 2C building block with the parameter $\theta$.}
    \label{fig:CC to 2CNOT}
\end{figure}

The two building blocks to be transformed are shown in figure \ref{fig:3C and 2C}. Then based on  $A\otimes B=\mathcal{M} U\mathcal{M}^{\dagger}$, 3C block and 2C block are mapped to $U_3(\theta/2,0,0)\otimes U_3(\theta/2,0,0)$ and $U_3(\theta/2,-\pi,\pi/2)\otimes U_3(\theta/2,-\pi,-\pi/2)$ respectively. Their elementary gate representations are $R_y(\theta/2)\otimes R_y(\theta/2)$ and $R^{\dagger}_y(\theta/2)S^{\dagger}\otimes R^{\dagger}_y(\theta/2)S$. Finally the optimal circuits are obtained based on $U=\mathcal{M}^{\dagger} (A\otimes B)\mathcal{M}$ as shown in figures \ref{fig:CCC to 2CNOT} and \ref{fig:CC to 2CNOT}.

\begin{figure}[h]
    \centerline{
     \Qcircuit @C = 0.8em @R = 0.5em  @!R{
     &\gate{S} &\gate{H} &\ctrl{1} &\gate{R_y(\theta/2)} &\ctrl{1} &\gate{H} &\gate{S^{\dagger}} &\qw\\\
     &\gate{S} &\qw      &\targ    &\gate{R_y(\theta/2)} &\targ    &\qw      &\gate{S^{\dagger}} &\qw
     }}
    \caption{The optimal circuit for 3C building block with the parameter $\theta$.}
    \label{fig:CCC to 2CNOT}
\end{figure}

Another common method is the approximate compiling method that attempts to find the best approximate circuit \emph{V} for the target \emph{U} up to a global phase by minimizing the cost \cite{Khatri2019}, e.g. the Frobenius norm $\frac{1}{2}\vert\vert V-U\vert\vert^2_F$. And the global phase introduces no impact to the evaluation of the Hamiltonian expectation, even in the circuit cutting scenario when all the elementary gates of each compiled $V$ are partitioned into the same fragment. Therefore this strategy can be universally applied to both ansatze.

\section{Complexity Analysis of the Ansatze}
\label{appen: building block complexity analysis}
\renewcommand{\thefigure}{C\arabic{figure}}
\renewcommand{\thetable}{C\arabic{table}}
\renewcommand{\theequation}{C\arabic{equation}}
The number of 3C blocks (and parameters) used for constructing CCC ansatz
accumulates as follows:\\
\begin{table}[h!]
    \centering
    \begin{threeparttable}
    \scalebox{0.7}{
    \renewcommand\arraystretch{2}{
    \begin{tabular}{c c c c c c}
    Layer             &1      &2        &3     &\ldots      &$k$                    \\
    $i$th qubit         &1      &3        &5     &\ldots      &$2k-1$               \\
    \# of 3Cs  &$n-k$    &$n-k-1$    &$n-k-2$ &\ldots      &$n-2k+1$   \\ % \hline\hline
    \end{tabular}}}
    \end{threeparttable}
\end{table}

So the total number is the sum of the values in the arithmetic sequence above, calculated as $k(n-k)-\frac{k(0+k-1)}{2}=nk-\frac{3k^2}{2}+\frac{k}{2}$. Then when $k=\frac{n}{3}$, we can obtain roughly the highest complexity $\frac{n^2}{6}$.

The number 2C blocks used for constructing CC ansatz can be evaluated as follows with odd $k$:\\

\begin{table}[h!]
    \centering
    \begin{threeparttable}
    \scalebox{0.7}{
    \renewcommand\arraystretch{2}{
    \begin{tabular}{c c c c c c}
    % \hline\hline
    Layer             &1      &2        &3     &\ldots      &$\frac{k+1}{2}$                    \\
    $i$th qubit         &1      &3        &5     &\ldots      &$k$               \\
    \# of 2Cs  &$n-1$    &$n-3$    &$n-5$ &\ldots      &$n-k$   \\ % \hline\hline
    \end{tabular}}}
    \end{threeparttable}
\end{table}

So the total number is about $(\frac{k+1}{2})(\frac{n-1+n-k}{2})=\frac{n(k+1)}{2}-\frac{k^2}{4}-\frac{k}{2}-\frac{1}{4}$. The upper bound can be loosely set to $\frac{n^2}{4}$ when $k=\frac{n}{2}$. {As can be seen, in the cases with large $n$ and small $k$, the number of blocks used in CC ansatz is almost half that of CCC ansatz.}

\section{Reduction of ``Measure and Prepare'' Channels} %IV IV IV IV IV IV
\label{appen: XY reduction}
\renewcommand{\thefigure}{D\arabic{figure}}
\renewcommand{\thetable}{D\arabic{table}}
\renewcommand{\theequation}{D\arabic{equation}}
Reducing the sampling complexity is a major research direction for the practicality of circuit cutting technique \cite{Perlin2021,Lowe2023}. During the evolution of a quantum circuit, the invalid states will be cancelled out based on the destructive inference phenomena. However, the circuit cutting technique distributes these redundant invalid components into different ``measure-and-prepare'' channels, which prevents the cancellation. This is the reason that the sampling complexity of the circuit cutting technique is exponential with respect to the total number of cut qubits. If the invalid states in different channels can be reduced in some way, the bulgy sampling complexity can slim down as well. Surprisingly, we found that all the redundancies can be eliminated when the number of cut qubit between the adjacent fragments is 1. The sampling complexity reduces from $O(2^p)$ to $O(1)$, with $p$ the total number of cut qubits.

\begin{figure}[ht]
    \centerline{
    \Qcircuit @C = 0.8em @R = 0.5em  @!R{
     &\lstick{\vert0\rangle} &\multigate{1}{U}  &\qw               &\qw                &\qw              &\qw              &\qw              &\qw        \\\
     &\lstick{\vert0\rangle} &\ghost{U}         &\multigate{1}{U}  &\qw                &\qw              &\qw              &\qw              &\qw        \\\
     &\lstick{\vert0\rangle} &\qw               &\ghost{U}        \ar@{.}[]+<1.4em,0.8em>;[d]+<1.4em,0.6em> &\multigate{1}{U}   &\qw              &\qw              &\qw              &\qw        \\\
     &\lstick{\vert0\rangle} &\qw               &\qw               &\ghost{U}          &\multigate{1}{U} &\qw              &\qw              &\qw        \\\
     &\lstick{\vert0\rangle} &\qw               &\qw               &\qw                &\ghost{U}   \ar@{.}[]+<1.4em,0.8em>;[d]+<1.4em,0.6em>     &\multigate{1}{U} &\qw              &\qw        \\\
     &\lstick{\vert0\rangle} &\qw               &\qw               &\qw                &\qw              &\ghost{U}        &\multigate{1}{U}  &\qw       \\\
     &\lstick{\vert0\rangle} &\qw               &\qw               &\qw                &\qw              &\qw              &\ghost{U}         &\qw       \\\
    }}
    \caption{Illustration of a single-layer staircase structure circuit.}
    \label{fig:single layer of staircase}
\end{figure}
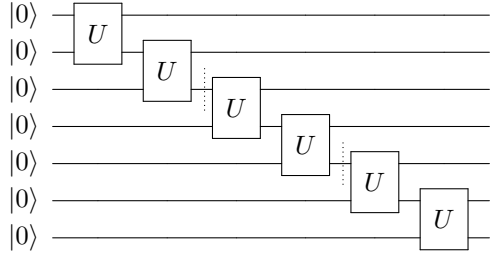

\begin{figure}[ht]
\centering
\begin{subfigure}{0.25\textwidth}
     \centering
     \Qcircuit @C = 1em @R = 0.6em{
&\gate{H}     &\meter   &  & & &\lstick{\vert j\rangle\langle j\vert}   &\gate{H} &\qw \\
\quad
}
     \caption{}
    \end{subfigure}

\begin{subfigure}{0.35\textwidth}
     \centering
     \Qcircuit @C = 1em @R = 0.6em{
&\gate{S^\dagger} &\gate{H}    &\meter   & & &  &\lstick{\vert j\rangle\langle j\vert}   &\gate{H} &\gate{S} &\qw \\
\quad
}
     \caption{}
    \end{subfigure}
\caption{The decomposition of ``measure-and-prepare'' channels of (a) \emph{X}, (b) \emph{Y} basis. State $\vert j\rangle$ with $j\in\{0, 1\}$ is the measurement result.}
\label{fig:XY measure and prepare channels}
\end{figure}
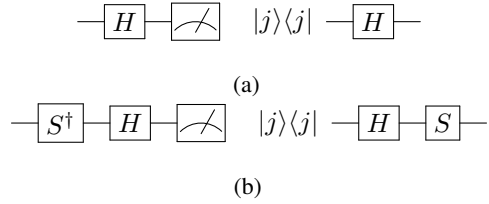

An illustrative circuit with a single layer of staircase structure is shown in figure \ref{fig:single layer of staircase}. The $NOT$ gates used for choosing specific column are synthesized into two-qubit $U$ gates. Remember that the amplitudes of the Dicke states prepared by our ansatze are real numbers. The dotted lines denote the cut position. Consequently the circuit is partitioned into 3 fragments. The first fragment is in fact also a single layer of staircase structure circuit. Then for CCC ansatz the state prepared by this fragment is a Dicke state with a unique $k$. (I) If the value of the cut qubit is 0, then the values of the uncut qubits are apparently different because they correspond to different basis states with the same Hamming weight $k$. (II) If the value of the cut qubit is 1, the uncut qubits are not equal to each other in the same way as when they have the unique Hamming weight $k-1$ now. (III) If the value of the cut qubit is a superposition of 0 and 1, then the uncut qubits are guaranteed to be different because basis states with different Hamming weights are different. Therefore, measuring the cut qubit in the $X$ basis, see figure \ref{fig:XY measure and prepare channels} (a), can be formalized as
\begin{equation}
\begin{split}
H\otimes I \vert \psi\rangle =& H\otimes I (\vert 0\rangle\sum_x\vert \psi^0_x\rangle + \vert 1\rangle\sum_y\vert \psi^1_y\rangle)   \\
=& \frac{1}{\sqrt{2}}(\vert 0\rangle + \vert 1\rangle)\sum_x\vert \psi^0_x\rangle + \frac{1}{\sqrt{2}}(\vert 0\rangle - \vert 1\rangle)\sum_y\vert \psi^1_y\rangle \\
=& \frac{1}{\sqrt{2}}\vert 0\rangle(\sum_x\vert \psi^0_x\rangle + \sum_y\vert \psi^1_y\rangle) + \frac{1}{\sqrt{2}}\vert 1\rangle(\sum_x\vert \psi^0_x\rangle - \sum_y\vert \psi^1_y\rangle),
\end{split}
\end{equation}
where $\sum_x\vert \psi^0_x\rangle$ denotes the basis states with Hamming weight $k$, while $\sum_y\vert \psi^1_y\rangle$ denotes the ones with Hamming weight $k-1$. It is quite obvious that no interference happens among these states. So we have the reconstruction term \cite{Tang2021}
\begin{equation}
\begin{split}
Tr(\rho X)&=Tr(\rho HZH)=Tr(H\rho HZ) \\
&=Tr(H\rho H(\vert 0\rangle\langle 0\vert-\vert 1\rangle\langle 1\vert)) \\
&=Tr(H\rho H\vert 0\rangle\langle 0\vert)-Tr(H\rho H\vert 1\rangle\langle 1\vert) \\
&=\langle 0\vert H\rho H\vert 0\rangle-\langle 1\vert H\rho H\vert 1\rangle \\
&=p(\vert 0\rangle\vert \psi^{\prime} \rangle)-p(\vert 1\rangle\vert \psi^{\prime}\rangle) \\
&=0,
\end{split}
\end{equation}
where $\vert \psi^{\prime}\rangle$ is the state of uncut qubits. That is, the measurement in the $X$ basis has no contribution to the reconstruction of useful information. A similar derivation for eliminating the $Y$ basis, see figure \ref{fig:XY measure and prepare channels} (b), is
\begin{equation}
\begin{split}
(HS^\dagger)\otimes I \vert \psi\rangle =& (HS^\dagger)\otimes I (\vert 0\rangle\sum_x\vert \psi^0_x\rangle + \vert 1\rangle\sum_y\vert \psi^1_y\rangle)   \\
=& \frac{1}{\sqrt{2}}\vert 0\rangle(\sum_x\vert \psi^0_x\rangle -i \sum_y\vert \psi^1_y\rangle) + \frac{1}{\sqrt{2}}\vert 1\rangle(\sum_x\vert \psi^0_x\rangle +i \sum_y\vert \psi^1_y\rangle),
\end{split}
\end{equation}
the similar result can be obtained as $Tr(\rho Y)=p(\vert 0\rangle\vert \psi^{\prime} \rangle)-p(\vert 1\rangle\vert \psi^{\prime}\rangle)=0$.

In consequence, the preparations of $H\vert j\rangle\langle j\vert$ and $SH\vert j\rangle\langle j\vert$ at the corresponding channel of the second fragment can be ignored, leaving $\vert 0\rangle$ and $\vert 1\rangle$ the only inputs. In this situation, the second fragment with input $\vert 0\rangle$ or $\vert 1\rangle$ is another Dicke state preparation circuit. In the iterative manner, the $X$ and $Y$ bases are eliminated in all fragments.

Ultimately the single qubit identity channel Id \cite{Peng2020} can be decomposed as
\begin{equation}
\operatorname{Id(\rho)} = \sum^4_{i=1}c_i{\rm Tr}(O_i\rho)\rho_i,
\end{equation}
where the Pauli observables $O_i$, the corresponding eigenprojectors $\rho_i$, and the eigenvalues $c_i$ remain
\begin{equation}
\begin{aligned}
O_1 &= I, &\rho_1 &= \vert 0\rangle\langle 0\vert, &c_1 &= +1/2, \\
O_2 &= I, &\rho_2 &= \vert 1\rangle\langle 1\vert, &c_2 &= +1/2, \\
O_3 &= Z, &\rho_3 &= \vert 0\rangle\langle 0\vert, &c_1 &= +1/2, \\
O_4 &= Z, &\rho_4 &= \vert 1\rangle\langle 1\vert, &c_2 &= -1/2. \\
\end{aligned}
\end{equation}

\noindent Further, the outputs of channels $I$ and $Z$ are the same, except the opposite sign in certain states. After cancelling these kind of states out, we found that the remaining states can be produced by observables $\vert 0\rangle\langle 0\vert$ and $\vert 1\rangle\langle 1\vert$ as
\begin{equation}
\begin{aligned}
\rho = \langle 0\vert\rho \vert 0\rangle \vert 0\rangle\langle 0\vert + \langle 1\vert\rho \vert 1\rangle \vert 1\rangle\langle 1\vert.
\label{eq:reduced I channel}
\end{aligned}
\end{equation}

\noindent This can be regarded as the sub-channels of channel $I$, which only communicate useful information. The other sub-channels that communicate redundant information are eliminated. No redundancy means there is no need to perform extra sampling to eliminate the invalid states. As a result, the sampling complexity decreases to $O(\epsilon^{-2})$ with $\epsilon$ the accuracy. This is very amenable for circuit cutting-based distributed quantum computing because the sampling complexity no longer relates to the total number of cut qubits any more, which guarantees higher fidelity and lower variance of the result when cutting the original circuit into more fragments. This fits perfectly with our original idea of finding a larger-scale application suitable for the current NISQ computers. The result above inspires us that the identity channel Id should be able to be decomposable into fewer sub-channels when the state prepared by the ansatz possesses some form of symmetry or non-exponential degree of freedom.

CC ansatz is a little more complex because there are extra basis states that exist when $k$ belongs to interval $[3,n-3]$. Fortunately we found that the Hamming weights of the extra basis states are all $k-2i$ with positive integer $i$. This observation assures the validity of the conclusion for CCC ansatz, as the differences in the Hamming weights of their uncut qubits are at least 1. That is to say, the minimal difference between the nearest two Hamming weights of $k$ and $k-2$ is 1 when the values of the cut qubit are 1 and 0 respectively. In fact, the present result still holds when choosing multiple columns simultaneously, as the Hamming weights of the states prepared by this kind of subcircuits follow the above analysis.

\section{Experimental Schemes}
\label{appen: Experiment schemes and complexity}
\renewcommand{\thefigure}{E\arabic{figure}}
\renewcommand{\thetable}{E\arabic{table}}
\renewcommand{\theequation}{E\arabic{equation}}
For the cases of $k=2,3$ in CCC ansatz, the number of independent columns are $\lfloor\frac{n}{2}\rfloor$ and $\lfloor\frac{n-1}{2}\rfloor$ respectively. Here ``independent'' means any target basis state can only appear in a single column.  In other words, there is no overlap between any two subspaces spanned by these subcircuits. For $k=2$, we find that these are columns 3, 12, 48,\ldots, $3\times4^{\lfloor\frac{n}{2}\rfloor-1}$. Each column can be prepared by the column chosen circuit $X^{\otimes 2}$ operating on the corresponding qubits, e.g. column 3 corresponds to $X^{\otimes 2}$ operating on the lowest 2 qubits. For $k=3$, these are columns  $3+4^{\lceil\frac{n}{2}\rceil-1}$, $12+4^{\lceil\frac{n}{2}\rceil-1}$, $48+4^{\lceil\frac{n}{2}\rceil-1}$,\ldots, $3\times4^{\lfloor\frac{n-1}{2}\rfloor-1}+4^{\lceil\frac{n}{2}\rceil-1}$. Each column can be prepared by $X^{\otimes 2}$ operating on the corresponding qubits and an $X$ gate on qubit $2(\lceil\frac{n}{2}\rceil-1)$ in the meantime.

For CC ansatz, similar distribution pattern can be found. In $k=2$ case, the ansatz is split into $\lfloor\frac{n}{2}\rfloor$ subcircuits, each of which corresponds to a pair of columns that are ($2^{n-3}\times3$, $2^{n-3}\times1$), ($2^{n-5}\times3$, $2^{n-5}\times1$), \ldots, and the last pair of columns is ($2^0\times3$, $2^0\times1$) when $n$ is odd or the last column is 1 when $n$ is even. Each pair of columns can be prepared by $R_y\otimes X$ operating on the corresponding qubits. The number of subcircuits is $\lfloor\frac{n-1}{2}\rfloor$ for the case of $k=3$. The columns pair are ($2^{n-1}+2^{n-4}\times3$, $2^{n-1}+2^{n-4}\times1$), ($2^{n-1}+2^{n-6}\times3$, $2^{n-1}+2^{n-6}\times1$), \ldots, and the last pair of columns is ($2^{n-1}+2^0\times3$, $2^{n-1}+2^0\times1$) when $n$ is even or the last column is 1 when $n$ is odd. Each pair of columns can be prepared by $R_y\otimes X$ operating on the corresponding qubits and an $X$ gate on the highest qubit in the meantime.

\begin{table}[ht] 
    \setlength{\abovecaptionskip}{0.1cm}
    \setlength{\belowcaptionskip}{0.1cm}
    \caption{Complexity of HDC scheme for cases of $k=2,3$ for both ansatze.}
    \label{tab:Complexity of HDA}
    \centering
    \begin{threeparttable}
    \scalebox{0.87}{
    \renewcommand\arraystretch{2}{
    \begin{tabular}{c c c c}
    \hline\hline
    \multicolumn{2}{c}{}                    &\# of subcircuits           &\# of Paras per sub                             \\ \hline
    \multirow{2}{*}{CCC ansatz}   &$k=2$  &$\lfloor\frac{n}{2}\rfloor$ &$2i$, $i\in [\lfloor\frac{n}{2}\rfloor]$\tnote{1} \\ 
                                    &$k=3$  &$\lfloor\frac{n-1}{2}\rfloor$ &$n-1$                                         \\ \hline
    \multirow{2}{*}{CC ansatz}  &$k=2$  &$\lfloor\frac{n}{2}\rfloor$ &$2i$, $i\in [\lfloor\frac{n}{2}\rfloor]$            \\ 
                                    &$k=3$  &$\lfloor\frac{n-1}{2}\rfloor$ &$n$                                           \\ \hline\hline
    \multicolumn{4}{l}{
    \begin{minipage}{8cm}
     $^1$ $[\lfloor\frac{n}{2}\rfloor]$ refers to integers in interval $[1, \lfloor\frac{n}{2}\rfloor]$.
    \end{minipage}
    }\\
    \end{tabular}}}
    \end{threeparttable}
\end{table}

\begin{figure}[ht]
\centering
\begin{subfigure}{0.4\textwidth}
     \centering
     \Qcircuit @C = 0.4em @R = 0.3em @!R{
     &\lstick{\vert0\rangle} &\qw       &\targ     &\ctrl{1}   &\targ      &\qw       &\qw        &\qw        &\qw       &\qw        &\qw       &\qw       &\qw        &\qw       &\qw \\\
     &\lstick{\vert0\rangle} &\qw       &\ctrl{-1} &\gate{R_y} &\ctrl{-1}  &\targ     &\ctrl{1}   &\targ      &\qw       &\qw        &\qw       &\qw       &\qw        &\qw       &\qw \\\
     &\lstick{\vert0\rangle} &\qw       &\qw       &\qw        &\qw        &\ctrl{-1} &\gate{R_y} &\ctrl{-1}  &\targ     &\ctrl{1}   &\targ     &\qw       &\qw        &\qw       &\qw \\\
     &\lstick{\vert0\rangle} &\gate{X}  &\qw       &\qw        &\qw        &\qw       &\qw        &\qw        &\ctrl{-1} &\gate{R_y} &\ctrl{-1} &\targ     &\ctrl{1}   &\targ     &\qw \\\
     &\lstick{\vert0\rangle} &\gate{X}  &\qw       &\qw        &\qw        &\qw       &\qw        &\qw        &\qw       &\qw        &\qw       &\ctrl{-1} &\gate{R_y} &\ctrl{-1} &\qw \\\ \gategroup{1}{4}{3}{9}{1.4em}{--}
      \quad
}
     \caption{}
    \end{subfigure}

\begin{subfigure}{0.4\textwidth}
     \centering
     \Qcircuit @C = 0.4em @R = 0.3em @!R{
     &\lstick{\vert0\rangle} &\qw       &\targ     &\ctrl{1}   &\targ      &\qw       &\qw        &\qw        &\qw       &\qw        &\qw       &\qw       &\qw        &\qw       &\qw \\\ 
     &\lstick{\vert0\rangle} &\gate{X}  &\ctrl{-1} &\gate{R_y} &\ctrl{-1}  &\targ     &\ctrl{1}   &\targ      &\qw       &\qw        &\qw       &\qw       &\qw        &\qw       &\qw \\\
     &\lstick{\vert0\rangle} &\gate{X}  &\qw       &\qw        &\qw        &\ctrl{-1} &\gate{R_y} &\ctrl{-1}  &\targ     &\ctrl{1}   &\targ     &\qw       &\qw        &\qw       &\qw \\\
     &\lstick{\vert0\rangle} &\qw       &\qw       &\qw        &\qw        &\qw       &\qw        &\qw        &\ctrl{-1} &\gate{R_y} &\ctrl{-1} &\targ     &\ctrl{1}   &\targ     &\qw \\\
     &\lstick{\vert0\rangle} &\qw       &\qw       &\qw        &\qw        &\qw       &\qw        &\qw        &\qw       &\qw        &\qw       &\ctrl{-1} &\gate{R_y} &\ctrl{-1} &\qw \\\
      \quad
}
     \caption{}
    \end{subfigure}    

    \caption{The two subcircuits for, (a) column 3 and (b) column 12, of $\vert D^5_2\rangle$ for the first hybrid distributed ansatz. The two blocks in the dashed box are eliminated. Actually, the subspace spanned by column 3 is contained in the reverse symmetric subspace spanned by column 12. Hence we can reverse the subspaces with less qubits to reduce the difficulty of executing larger-qubit subcircuits.}
    \label{fig:two subcircuits of the example}
\end{figure}
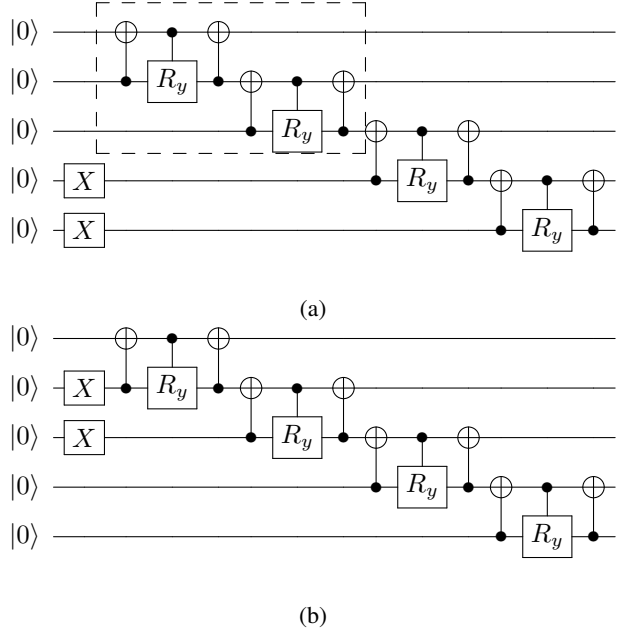

The above results can be verified easily via the matrix representation of $U_n$, see figure \ref{fig:Unmatrices} in appendix \ref{appen: theoretical support}. Without splitting classically the ansatze into subcircuits, the number of parameters of each ansatz is $O(nk)$ and the sampling complexity is $p16^{O(p)}$ when the number of fragments is $p+1$. Commonly the total number of cut qubits $p$ should be linear in $n$ with the constant factor small enough to guarantee the advantage of quantum computing. Differently, here $p$ has no impact on the total sampling complexity $O(\epsilon^{-2})$ of each subcircuit explicitly. In the meantime, the errors can be mitigated more effectively by measurement error mitigation assisted by the symmetry property, see section \ref{sec:experi} in the main text. In table \ref{tab:Complexity of HDA}, we provide the detailed complexities of HDC for our ansatze with $k=2$ and $k=3$. For cases of $k=2$, the building blocks in high qubits of the subcircuits with small column indices can be reduced because they have no impact on the output. Therefore the number of parameters for $k=2$ is much smaller. As an example, figure \ref{fig:two subcircuits of the example} shows the two subcircuits of the hybrid distributed $\vert D^5_2\rangle$ state of CCC ansatz. The subcircuit corresponding to column 3 (shown in figure \ref{fig:two subcircuits of the example} (a)) utilizes just 3 qubits and 2 parameters. Clearly, the total sampling complexity decreases dramatically.

Without doubt, multiple columns can be prepared simultaneously to reduce the number of subcircuits. For more practical cases with much larger $k$, the number of independent columns increases in a superlinear manner in $n$. Hence we need to select multiple columns that overlap with each other, or import extra states, to obtain an affordable number of subcircuits that can provide quantum advantages, see appendix \ref{appen: theoretical support}.

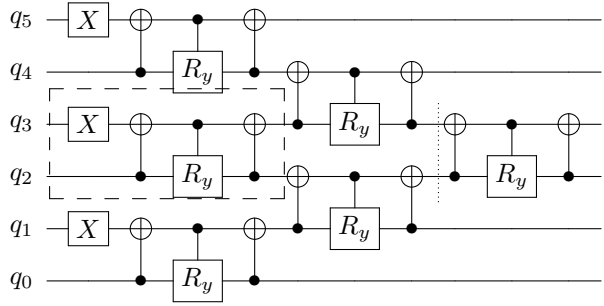
\begin{figure}[ht]
    \centerline{
    \Qcircuit @C = 0.8em @R = 0.4em  @!R{
     &\lstick{q_5} &\gate{X}  &\targ     &\ctrl{1}   &\targ      &\qw       &\qw        &\qw        &\qw       &\qw        &\qw       &\qw \\\
     &\lstick{q_4} &\qw       &\ctrl{-1} &\gate{R_y} &\ctrl{-1}  &\targ     &\ctrl{1}   &\targ      &\qw       &\qw        &\qw       &\qw \\\
     &\lstick{q_3} &\gate{X}  &\targ     &\ctrl{1}   &\targ      &\ctrl{-1} &\gate{R_y} &\ctrl{-1} \ar@{.}[]+<1em,0.8em>;[d]+<1em,1em>   &\targ     &\ctrl{1}   &\targ  &\qw \\\
     &\lstick{q_2} &\qw       &\ctrl{-1} &\gate{R_y} &\ctrl{-1}  &\targ     &\ctrl{1}   &\targ     \ar@{.}[]+<1em,0.8em>;[d]+<1em,1em>   &\ctrl{-1} &\gate{R_y} &\ctrl{-1} &\qw \\\
     &\lstick{q_1} &\gate{X}  &\targ     &\ctrl{1}   &\targ      &\ctrl{-1} &\gate{R_y} &\ctrl{-1}  &\qw       &\qw        &\qw        &\qw \\\ 
     &\lstick{q_0} &\qw       &\ctrl{-1} &\gate{R_y} &\ctrl{-1}  &\qw       &\qw        &\qw        &\qw       &\qw        &\qw        &\qw \\\  \gategroup{3}{3}{4}{6}{1.4em}{--}
    }}
    \caption{Illustration of HDC design of $\vert D^6_3\rangle$. The entangled part shown in the dashed box is replaced by product states $\vert 10 \rangle$ and $\vert 01 \rangle$. The entanglement in the last layer (layer of $U_n$) should be reserved.}
    \label{fig:HDC D6_3}
\end{figure}

In order to show the high scalability and flexibility of our HDC scheme for solving combinatorial optimization problems, here we provide a more practical scheme for preparing the widely used Dicke state $\vert D^n_{\lfloor n/2\rfloor}\rangle$ which also reserves the identity channel Id of equation \eqref{eq:reduced I channel}, see appendix \ref{appen: XY reduction}. We directly give the result: The Dicke state $\vert D^n_{\lfloor n/2\rfloor}\rangle$ prepared by CCC ansatz can be split classically into $\lceil n/2\rceil - 1$ subcircuits, each of which can be cut evenly into 2 fragments by cutting only two qubits. And the remaining 3C block can be evaluated trivially in a classical manner.

Its complexity is much lower than that of directly applying circuit cutting technique. As mentioned above, the basis states prepared by these subcircuits have overlap. In this case, we can solve a problem related to Dicke states $\vert D^n_{\lfloor n/2\rfloor}\rangle$ by a $n/2$-qubit NISQ computer. For example, the Dicke state $\vert D^{60}_{30}\rangle$ can be split into 29 subcircuits, which has potential to achieve quantum advantage. This result is obtained by equivalently replacing the entangled part between the two fragments with combinations of correspondingly multiple product circuits. Specifically, for Dicke state $\vert D^6_3\rangle$, see figure \ref{fig:HDC D6_3}, to cut the ansatz evenly, we choose to split classically the superposition state of $\vert 01\rangle$ and $\vert 10\rangle$ prepared by the circuit in the dashed box into two Kronecker product parts and then cut off the additional 3C block after the dotted line on qubits $q_2$ and $q_3$. This partition scheme also promises the 2 fragments in each subcircuit can always produce all basis states with the same Hamming weight respectively which can be used for high-quality error mitigation.

Our numerical simulations show that, to obtain the optimal solution in the case $\vert D^{40}_{20}\rangle$, the classical brute-force takes over ten days while the HDC scheme only takes several hours. The simulation of the 20-qubit fragments consumes most of the time. A personal computer with Intel(R) Core(TM) i5-10500 CPU @ 3.10GHz and 32GB RAM (2666MHz 16G × 2) is used.

\section{Symmetric Space Partition Scheme} %VI VI VI VI VI VI
\label{appen: symmetric space partition scheme}
\renewcommand{\thefigure}{F\arabic{figure}}
\renewcommand{\thetable}{F\arabic{table}}
\renewcommand{\theequation}{F\arabic{equation}}

\begin{figure}[ht]
    \centering
    \begin{subfigure}{0.35\textwidth}
     \centering
     \includegraphics[width=\textwidth]{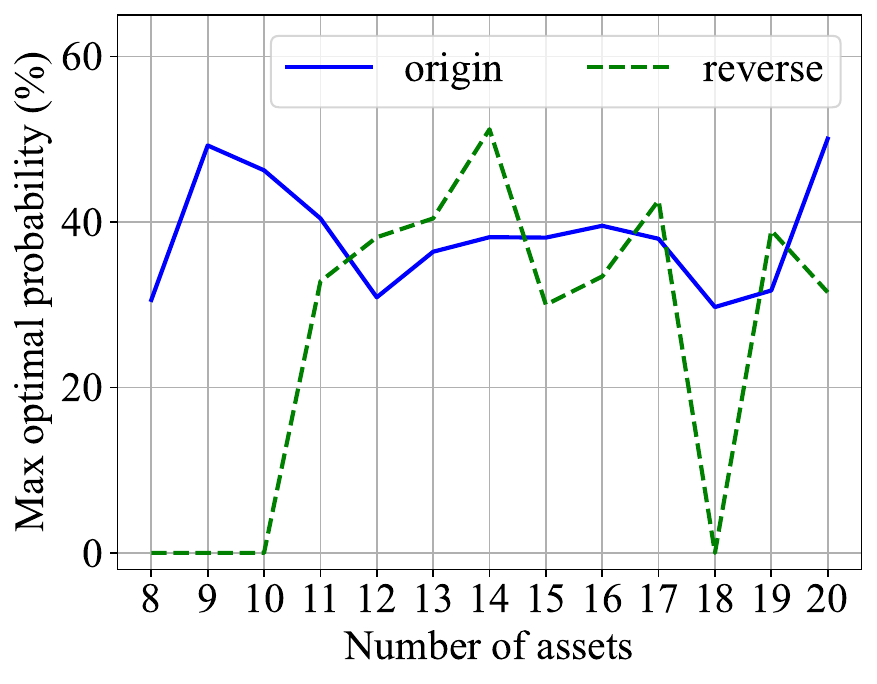}
     \caption{}
    \end{subfigure}
    \begin{subfigure}{0.35\textwidth}
     \centering
     \includegraphics[width=\textwidth]{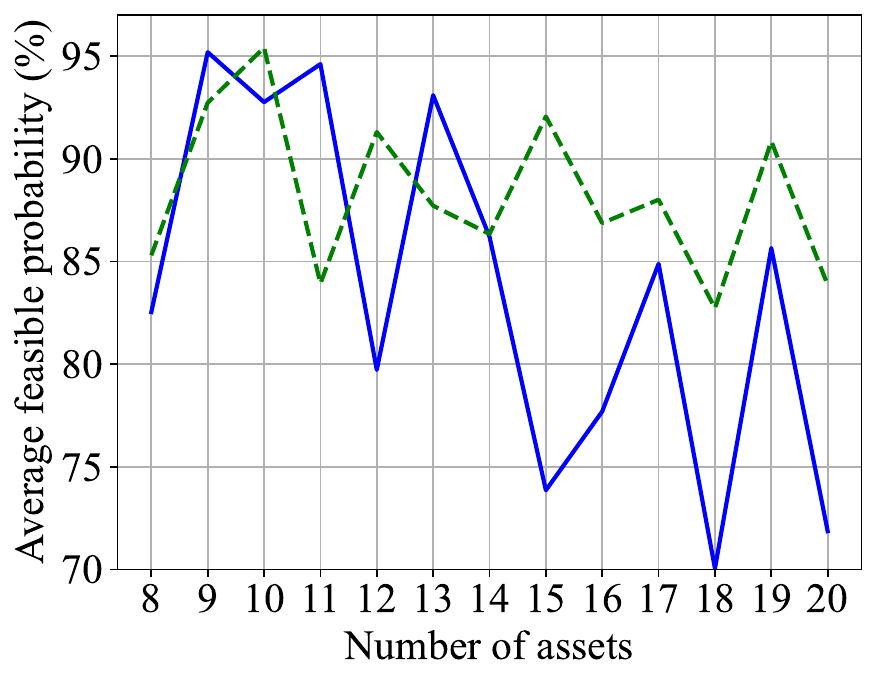}
     \caption{}
    \end{subfigure}
    \caption{The maximal/average probability of obtaining (a) the optimal solution and (b) the feasible solution up to 20 assets.}
    \label{fig:symmetric space partition scheme}
\end{figure}

According to the reverse symmetric property of Dicke state $\vert D^n_k\rangle$ mentioned in the main text, we can partition the space spanned by CC ansatz into two reverse symmetric subspaces, each of which contains one basis state of a symmetric pair that consists of two reverse symmetric basis states, such as 0011 and 1100, 1010 and 0101. Then we can employ one of the two subspaces as the search space, and perform the optimization process twice, corresponding to the assets in the original order and in the reversed order respectively. In practice, the perfectly bisected partition of these symmetric pairs is difficult to achieve. Hence we propose a compromised version that reduces non-targets to a large extent and preserves a reverse symmetric subspace, which is just required to remove the lowest $\lfloor (n-k)/2\rfloor +1$ 2C blocks in the staircase layer adjacent to $U_n$. This number has been confirmed through extensive numerical testing up to 29 qubits. In addition, a small number of basis states that are symmetrical, such as 10001, 01010, may not be contained in the search subspace which can be easily verified by directly calculating their expectations.

In this appendix, we also give the representative illustrations in figure \ref{fig:symmetric space partition scheme} of the symmetric space partition scheme. We performed a large number of numerical simulations and found that, in most of the experiments, both the subspaces spanned by the origin-order and reverse-order ansatze contain the optimal solution. Specifically to show the complementary characteristics of the two order cases in extreme cases, here the random seed 1213 for generating the financial data is chosen. The confidence level $\alpha$ is 0.25, and the budget is 4. And the simulation is performed with 5 different initializations of the parameters $\bm{\theta}$ in order to reduce the influence of bad parameter initialization, which could lead to the optimization converging to a result that contains the optimal solution with a tiny, even 0, probability. As illustrated in the figure, all pairs of ordered assets obtain the optimal solution with a sufficiently high probability. And the average probability of obtaining feasible solutions is always greater than 70\%.

\section{Performance Comparison of Different Layers}
\label{appen: multi layers of Un for CCC ansatz}
\renewcommand{\thefigure}{G\arabic{figure}}
\renewcommand{\thetable}{G\arabic{table}}
\renewcommand{\theequation}{G\arabic{equation}}

\begin{figure}[ht!]
    \centering
    \begin{subfigure}{0.45\textwidth}
     \centering
     \includegraphics[width=\textwidth]{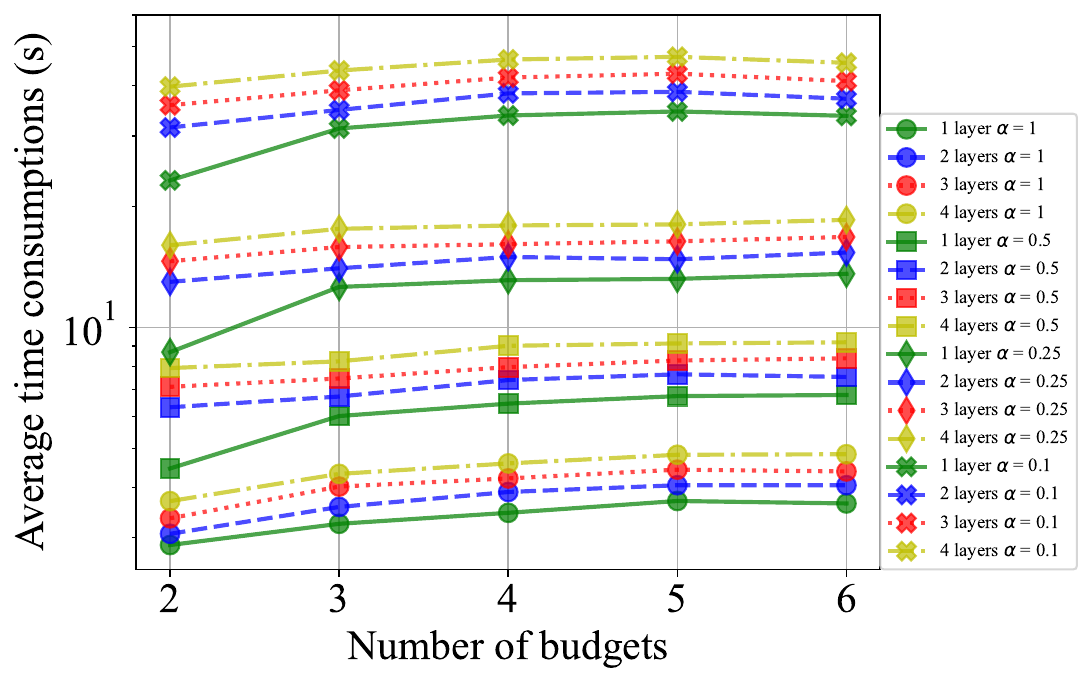}
     \caption{}
    \end{subfigure}
    
    \begin{subfigure}{0.4\textwidth}
     \centering
     \includegraphics[width=\textwidth]{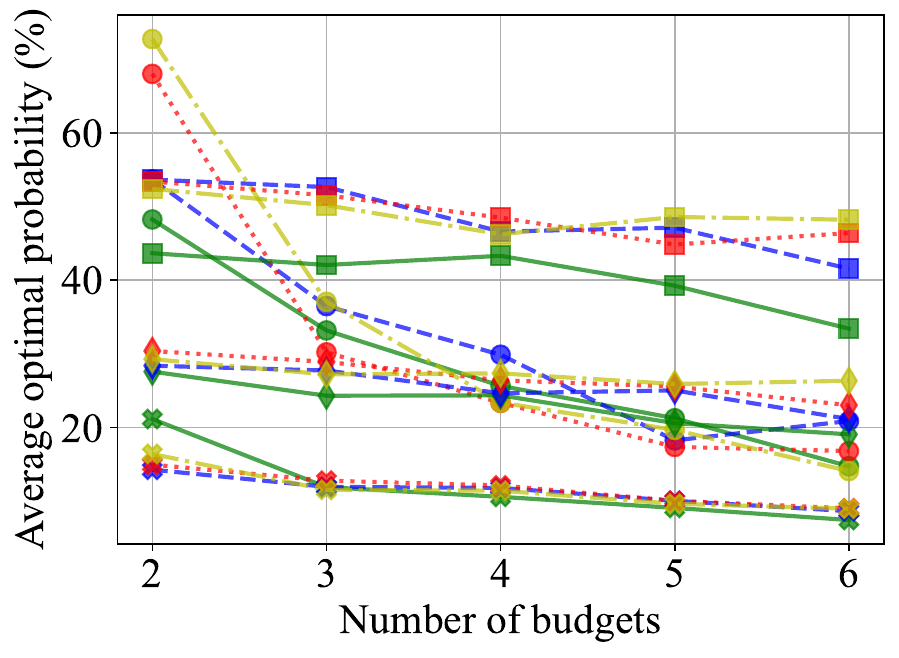}
     \caption{}
    \end{subfigure}
    \begin{subfigure}{0.4\textwidth}
     \centering
     \includegraphics[width=\textwidth]{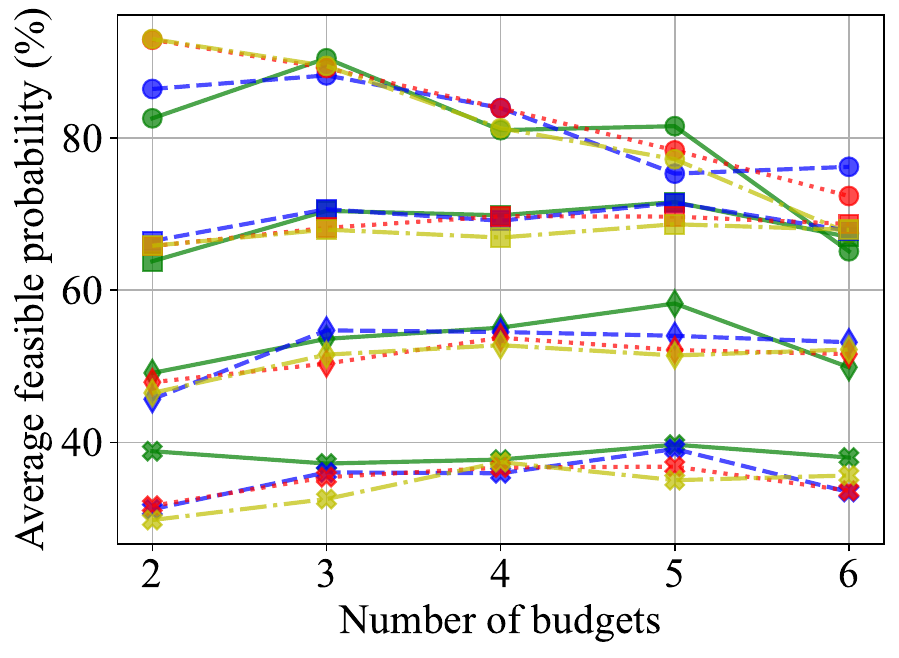}
     \caption{}
    \end{subfigure}
    \caption{Illustrative comparison of the performance of CCC ansatz with different layers of $U_n$ upon different confidence levels.}
    \label{fig:multi layers of Un for CCC ansatz}
\end{figure}

Theoretically, more parameters enhances the expressibility of the ansatz which is beneficial for obtaining the optimal solution. However, in the NISQ era, more parameters seems not to be an efficient strategy. More parameters leads to an ansatz with more gates and depth, which induces more errors. The accumulation of these errors in the calculation of each expectation upon the corresponding parameter configuration during the optimization process introduces more serious instability, which hinders the selection of the optimization direction.

The performance of CCC ansatz with 1 to 4 layers of $U_n$ is illustrated in figure \ref{fig:multi layers of Un for CCC ansatz}. The number of assets is set to 12, and the asset pool is randomly initialized 50 times for each budget. The results indicate, on average, as the number of layers increases, the time consumption increases on the corresponding confidence level as expected. However the probabilities of obtaining the optimal and feasible solutions are almost the same for slightly smaller $\alpha$. It is intuitive that the confidence level $\alpha$ should be reduced to the corresponding threshold for much bigger $n$ and $k$. As can be seen, the CVaR cost function with a small enough confidence level can guarantee the performance. Hence one layer of $U_n$ with a small $\alpha$ seems to be a much better choice for solving combinatorial optimization problems, not only for the NISQ era. 

\section{Experimental Setup}
\label{appen: experimental setup}
\renewcommand{\thefigure}{H\arabic{figure}}
\renewcommand{\thetable}{H\arabic{table}}
\renewcommand{\theequation}{H\arabic{equation}}

\begin{figure}[ht]
    \centering
    \includegraphics[width=0.35\textwidth]{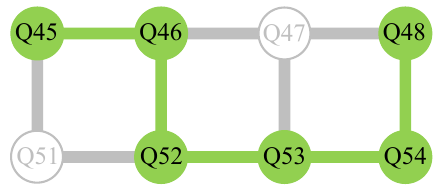}
    \caption{The topology of the 6 used qubits.}
    \label{fig:topology of qubits}
\end{figure}

The configurations of the two experiments are shown in table \ref{tab:Setup of experiments}. The qubit performance and topology are provided in table \ref{tab:Qubit performances} and figure \ref{fig:topology of qubits} respectively. $f_{00}$ ($f_{11}$) is the fidelity of measuring $\vert 0\rangle$ ($\vert 1\rangle$) when the true state is $\vert 0\rangle$ ($\vert 1\rangle$) .

\begin{table}[ht]
    \setlength{\abovecaptionskip}{0.2cm}
    \setlength{\belowcaptionskip}{0.2cm}
    \caption{Setup of simulations and experiments.}
    \label{tab:Setup of experiments}
    \centering
    \begin{threeparttable}
    \scalebox{0.7}{
    \renewcommand\arraystretch{2}{
    \begin{tabular}{c | c c c c c c | c}
    \hline\hline
    Number of assets                 &5      &15      &25       &35       &45       &55     &12      \\ \hline
    IvCL\tnote{1}                    &0.2    &0.1     &0.05     &0.05     &0.025    &0.025  &0.0125   \\ \hline
    Shots in ICI \tnote{2}           &8000   &4000    &2000     &2000     &1000     &1000   &3000     \\ \hline\hline
    \multicolumn{8}{l}{
    \begin{minipage}{8cm}
    $^1$ the Initial value of Confidence Level $\alpha$.\\
    $^2$ the Initial Confidence Interval $(0, \alpha]$.
    \end{minipage}
    }\\
    \end{tabular}}}
    \end{threeparttable}
\end{table}

\begin{table}[ht]
    \setlength{\abovecaptionskip}{0.2cm}
    \setlength{\belowcaptionskip}{0.2cm}
    \caption{Qubit performance.}
    \label{tab:Qubit performances}
    \centering
    \begin{threeparttable}
    \scalebox{0.7}{
    \renewcommand\arraystretch{2}{
    \begin{tabular}{c c c c c c q}
    \hline\hline
    Qubit         &$T_1$($\mu$s)      &$T_2$(ns)       &$f_{00}$($\%$)     &$f_{11}$($\%$)      &$f_{SQ}$($\%$)\tnote{1}       &\multicolumn{1}{l}{$f_{CZ}$($\%$)}    \\ \hline
    Q45           &26.6     &354       &97.3          &93.1          &99.75            &97.31\ \,     \\
    Q46           &23.8     &707       &95.1          &92.4          &99.61            &97.11\ \,     \\
    Q52           &19.3     &977       &97.0          &92.6          &99.81            &98.16\ \,    \\
    Q53           &31.4     &629       &97.2          &94.1          &99.82            &98.00\ \,     \\
    Q54           &22.7     &5941      &91.7          &85.4          &99.91            &97.68\ \,     \\
    Q48           &13.6     &339       &91.0          &83.2          &99.69            &\ \,  \\ \hline\hline
    \multicolumn{7}{l}{
    \begin{minipage}{6.5cm}
     $^1$ SQ represents single-qubit gate.
    \end{minipage}
    }\\
    \end{tabular}}}
    
    \end{threeparttable}
\end{table}

\end{appendices}
\printbibliography[title={References}]
\end{document}